\documentclass{emulateapj}

\shorttitle{Long-Slit Spectra and the Cusp/Core Problem}
\shortauthors{Spekkens, Giovanelli \& Haynes}

\begin{document}

\title{The Cusp/Core Problem in Galactic Halos: \\Long-Slit Spectra for a Large Dwarf Galaxy Sample}

\author{Kristine Spekkens, Riccardo Giovanelli\altaffilmark{1} \& Martha P. Haynes\altaffilmark{1}}
\affil{Center for Radiophysics and Space Research, Cornell University, Space Sciences Building, Ithaca, NY 14853}
\email{spekkens@astro.cornell.edu}

\altaffiltext{1}{National Astronomy and Ionosphere Center, Cornell University, 
Space Sciences Building, Ithaca, NY 14853. 
The National Astronomy and Ionosphere Center is operated by Cornell University 
under a cooperative agreement with the National Science Foundation.}

\submitted{Accepted for publication by the Astronomical Journal}

\begin{abstract}
We derive inner dark matter halo density profiles for a sample of 165 low-mass galaxies using rotation curves obtained from high-quality, long-slit optical spectra assuming minimal disks and spherical symmetry. 
For $\rho(r)~\sim~ r^{-\alpha}$ near the galaxy center we measure median inner slopes ranging from $\alpha_m = 0.22 \pm 0.08$ to $0.28 \pm 0.06$ for various subsamples of the data. This is similar to values found by other authors and in stark contrast to the intrinsic cusps ($\alpha_{int}\sim1$) predicted by simulations of halo assembly in cold dark matter (CDM) cosmologies. 
To elucidate the relationship between $\alpha_m$ and $\alpha_{int}$ in our data, we simulate long-slit observations of model galaxies with halo shapes broadly consistent with the CDM paradigm. 
Simulations with $\alpha_{int}=1/2$ and 1 recover both the observed distribution of $\alpha_m$ and correlations between $\alpha_m$ and primary observational parameters such as distance and disk inclination, whereas those with $\alpha_{int}=5/4$ are marginally consistent with the data. Conversely, the hypothesis that low-mass galaxies have $\alpha_{int}=3/2$ is rejected. While the simulations do not imply that the data favor intrinsic cusps over cores, they demonstrate that the discrepancy between $\alpha_m$ and $\alpha_{int}\sim1$ for our sample does not necessarily imply a genuine conflict between our results and CDM predictions: rather, the apparent cusp/core problem may be reconciled by considering the impact of observing and data processing techniques on rotation curves derived from long-slit spectra.
\end{abstract}


\section{Introduction: Long-Slit Spectra and the Cusp/Core Problem}
\label{intro}

 Within the standard cosmological cold dark matter (CDM) paradigm, structures form hierarchically via gravitational collapse of primordial density fluctuations. This framework has been successful at providing physical interpretations for a wide range of phenomena, notably the angular power spectrum of cosmic microwave background anisotropies, the large-scale structure of the galaxy distribution, and fundamental scaling relations in disk galaxies.

In recent years, significant theoretical progress has also been made in predicting halo shapes obtained from simulations of structure formation in a ($\Lambda$)CDM universe. Detailed collisionless numerical simulations of halo assembly 
(see Reed et al. 2003 and references therein)
 have led to a general consensus on the fundamental properties of the resulting CDM halos: their density distributions $\rho(r)$ deviate significantly from the power laws predicted by analytic calculations (e.g. Gunn \& Gott 1972); they have a broadly ``universal'' shape over many decades in mass that can be parameterized by simple fitting formulae; and they are {\it cuspy}, in that $\rho(r) \propto r^{-\alpha_{int}}$ at the smallest halo radii $r$ probed by simulations, and the intrinsic inner slope $\alpha_{int}$ is of order 1.

A simple analytic profile that encompasses these three characteristics was first proposed by Navarro et al. (1996; 1997, hereafter NFW) for dark matter halos:
\begin{equation}
\rho(r) = \frac{\rho_s}{(r/r_s)[1+(r/r_s)]^2}\,\,, 
\label{NFW}
\end{equation}
where $\rho_s$ and $r_s$ are the characteristic halo density and radius, respectively. 
Subsequent generations of simulations showed some disagreement over the value of $\alpha_{int}$ that best described the resulting halos: inner slopes shallower than the NFW value of $\alpha_{int}=1$  (Subramanian et al. 2000; Taylor \& Navarro 2001; Ricotti 2003) and as steep as $\alpha_{int}$=1.5 (e.g. Fukushige \& Makino 1997, 2001; Moore et al. 1998, 1999) have been proposed. A systematic study of these discrepancies has been undertaken by Navarro and collaborators, using a suite of high-resolution simulations that reliably probe galaxy-size halos down to small fractions of their virial radii (Hayashi et al. 2004; Navarro et al. 2003; Power et al. 2003). They find that while the density distribution does not converge to a well-defined asymptotic power-law at the innermost resolved radius, the average $\alpha_{int}$ is 1.1, 1.2, and 1.35 at 1\% of the virial radii of cluster, galaxy, and dwarf halos, respectively. The NFW slope $\alpha_{int}=1$ remains consistent with the simulated halo shapes, but cusps as steep as $\alpha_{int}=1.5$ are ruled out in almost all cases. Some impacts of the presence of baryons on these halo shapes have been investigated (e.g. Mo \& Mao 2004); however, fully self-consistent simulations of halo shapes including a variety of gasdynamical effects have not yet been achieved.

\begin{figure}
\epsscale{1.25}
\plotone{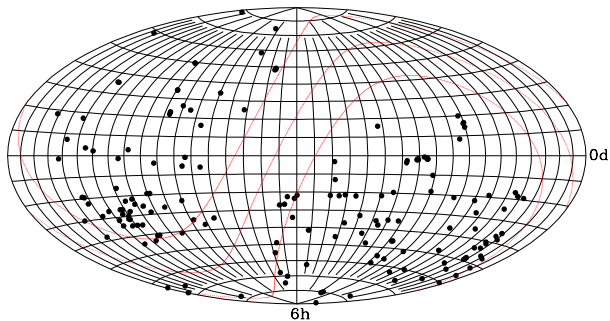}
\caption{Sky distribution of sample galaxies. Solid red lines denote the plane of the Milky Way ($b = 0^{\circ} \pm 20^{\circ}$).
\label{aitoff}}
\end{figure}

While collisionless halo shapes are consistent with the average properties of high redshift halos inferred from weak lensing studies (e.g. Hoekstra et al. 2004), on galaxy and cluster scales the agreement between theory and observations is much less certain. The majority of observations rely on luminous tracers embedded at small $r$ to probe the halo potential. Even though some studies of halo shapes prefer CDM profiles (e.g. Mahdavi \& Geller 2004;  Pointecouteau et al. 2004), a variety of evidence suggests that at least some real halos do not have the cusps predicted for the pre-baryon collapse systems, but rather {\it cores} in which the inner halo density has only a weak dependence on $r$ (or none at all). For instance, studies of gravitational arcs around clusters (Sand et al. 2004; but see El-Zant et al. 2004), models of gas flow along strong bars (Weiner et al. 2001), and the mass budget in the Milky Way from microlensing and stellar kinematics (Binney \& Evans 2001; Binney 2004) all yield measured inner halo slopes $\alpha_m$ that are much more ``corelike'' ($\alpha_{m} \lesssim 0.5$) than the cusps predicted by CDM. The rotation curves (RCs) of spiral galaxies have also been extensively used to constrain halo structure. In particular, Salucci (2001) concludes that halos have cores larger than the corresponding disk scale-lengths from a robust analysis of 137 disk-dominated systems that is immune to biases on the inner RC shape. This apparent conflict between CDM theory and observations in these and other systems has been called the ``cusp/core problem''. 

Among experiments that probe the structure of galactic halos, substantial resources have been devoted to deriving inner dark matter halo shapes for dwarf and low surface brightness (LSB) galaxies. Although in some systems the RC features correlate with those seen in the light distribution (Sancisi 2004), the global RC dynamics in dwarf and LSB galaxies are thought to be dark matter dominated even at small $r$ (e.g. Swaters 1999). The advantage in inferring their halo properties from observed gas dynamics is therefore twofold: first, there may be little contribution to the observed RC from baryonic components, in which case the RC shape is a direct manifestation of the inner halo structure. Second, baryonic collapse in the halo gravipotential well during disk formation may not significantly alter the halo structure. The halos of dwarf and LSB galaxies may therefore represent pristine, pre-baryon collapse distributions better suited to direct comparisons with collisionless CDM simulations than the halos of more massive or luminous systems.

Early observations of dwarf and LSB kinematics from \ion{H}{1} synthesis observations (e.g. Carignan \& Beaulieu 1989; C\^ot\'e et al. 1991; de Blok \& McGaugh 1997) yielded RCs that are well-described by models with cores rather than cusps. This apparent crisis for CDM (e.g. Flores \& Primack 1994; Moore 1994) has since been called into question, however, because beam smearing caused by limited spatial resolution in the inner halo systematically lowers the measured inner profile slopes (Blais-Ouellette et al. 1999, 2004; Swaters 1999; Swaters et al. 2000; van den Bosch et al. 2000; van den Bosch \& Swaters 2001; but see Gentile et al. 2004).  Alternatively, long-slit optical spectroscopy yields arcsecond-resolution RCs in relatively short integration times. There has therefore been a concerted effort to obtain long-slit spectra for many dwarf and LSB systems (e.g. Borriello \& Salucci 2001; de Blok et al. 2001, hereafter dB01; McGaugh et al. 2001; de Blok \& Bosma 2002; Marchesini et al. 2002; Swaters et al. 2003, hereafter S03).  A full characterization of the kinematics is then achieved by combining the resulting high resolution RCs with measurements of the outer RC shape in \ion{H}{1} (e.g. de Blok et al. 2001; Marchesini et al. 2002; S03; Gentile et al. 2004). Approximately 70 high-quality RCs have been obtained from these studies, and almost all are better described by cores than the cusps predicted by CDM. These data are clearly consistent with halos having cores rather than cusps; nonetheless, mass models of hybrid H$\alpha$+\ion{H}{1} RCs often cannot rule out the $\alpha_{int}=1$ case (S03, Hayashi et al. 2004). 

Many studies have focused on a simpler measure of $\alpha_m$ from long-slit spectra alone, in which the stellar mass is ignored and spherical symmetry is assumed (dB01; de Blok et al. 2003, hereafter dB03; S03).  This is clearly an over-simplification of the problem (e.g. Fall \& Efstathiou 1980; Salucci 2001), but one that produces meaningful results: minimum disks yield upper limits on the true steepness of the halo profile, and there is typically little change in the halo slope when a luminous component is introduced (dB01, de Blok \& McGaugh 1997). This technique, too, has revealed mostly core-like $\alpha_m$. However, even at arcsecond resolution a variety of observational uncertainties such as slit offsets, slit width and seeing, and geometric effects such as galaxy inclination all lead to an underestimate of $\alpha_m$ relative to the underlying $\alpha_{int}$. Moreover, inner halo shapes are measured in the region where RCs are most susceptible to folding errors, and where the relative uncertainties on the RC points are largest. As a result, raw long-slit spectra are heavily processed before the inner halo shapes are estimated, and this may also lead to a bias in the measured slopes. Indeed, the complexity of the exercise is illustrated by the significant differences in $\alpha_m$ sometimes inferred by different authors for the same galaxies, even when the same raw spectra are used (dB03; S03). 
To meaningfully compare the $\alpha_m$ obtained to the cuspy $\alpha_{int}$ predicted by CDM, then, a thorough assessment of these potential biases is required. 

However, there is no clear consensus among those who have undertaken this task. The simulations of dB03 indicate that no single systematic effect can reconcile their data with cuspy CDM halos, but those of S03 and Rhee et al. (2004) lead them to conclude that a range $0 \lesssim \alpha_{int} \lesssim 1$ is consistent with the $\alpha_m$ derived. With the present body of data, it is therefore unclear whether the halo shapes measured in this manner signal a genuine conflict for the CDM paradigm, or whether they can be reconciled by taking observing and data reduction techniques into account. 
 In the present paper we address this issue by deriving $\alpha_m$ for 165 low-mass galaxies under assumptions of spherical symmetry and a minimal disk, using techniques similar to those of dB01, dB03 and S03. We then perform simulations to determine the impact of the observing and data processing methods on our results. The size, homogeneity, and parameter space probed by the sample allow for both an independent measurement of the distribution of $\alpha_m$ in low-mass systems and a thorough investigation of the potential biases on the $\alpha_m$ values obtained.

The organization of this paper is as follows: our sample selection criteria are given in \S~\ref{sample}, and our analysis of the sample RCs and the values of $\alpha_m$ obtained are in \S~\ref{inversion}. The simulations we perform to compare the distribution of $\alpha_m$ with those expected for model galaxies with various $\alpha_{int}$ are described in \S~\ref{simulations}, and we compare our simulation results to the sample data in \S~\ref{results}. A summary of our findings as well as a list of caveats regarding the accurate determination of inner halo shapes is in \S6.


\begin{figure}
\epsscale{1.2}
\plotone{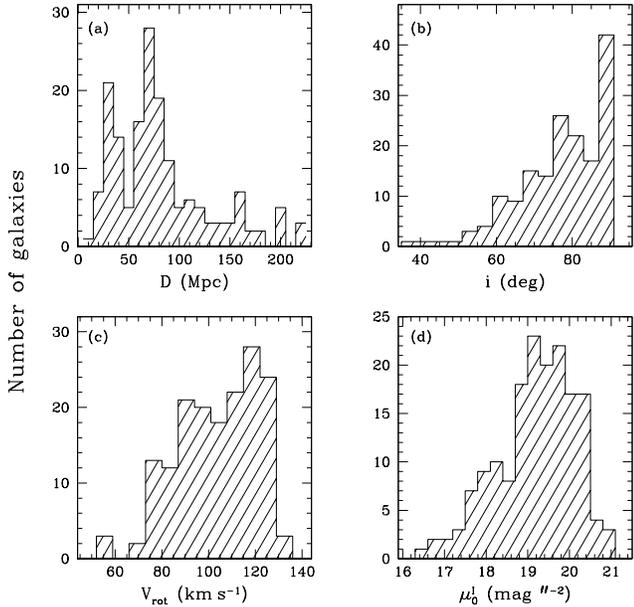}
\caption{Distribution of a) $D$, b) $i$, c) $V_{rot}$, and d) $\mu^I_0$ for the galaxy sample. 
\label{sam_properties}}
\end{figure}

\section{Sample Selection}
\label{sample}

The dwarf galaxy sample is selected from the SFI++ Tully-Fisher database
 maintained at Cornell University, a compilation of $\sim$4800 galaxies with accurate I-band photometry and estimates of rotational velocity $V_{rot}$ from either single-dish \ion{H}{1} or optical long-slit spectra. The SFI++ includes datasets published by our group (Dale \& Giovanelli 2000; Vogt et al. 2004 and references therein) and the southern spiral galaxy sample of Mathewson and collaborators (Mathewson et al. 1992; Mathewson \& Ford 1996) with photometry reprocessed by our group, as well as recently obtained data. For this study we select systems from the subset of galaxies in the SFI++ with archived H$\alpha$ RCs.

\begin{deluxetable*}{rrrcccccccccccc}
\tabletypesize{\scriptsize}
\tablecaption{Sample Properties and Measured Inner Halo Slopes \label{alpham}}
\tablewidth{0pt}
\tablehead{
\colhead{UGC/} & \colhead{$\alpha$ (J2000)} & \colhead{$\delta$ (J2000)} & \colhead{$V_{\odot}$} & \colhead{$D$} &
\colhead{$m_I$} & \colhead{$r_{opt}$} & \colhead{$\mu^I_0$} & \colhead{$i$} & \colhead{$V_{rot}$} & \colhead{$V_o$} &
\colhead{$r_{pe}$} & \colhead{$\beta$} & \colhead{$\alpha_m$} & \colhead{$\Delta \alpha_m$}\\
\colhead{AGC} & \colhead{($^{\rm{h}\,\, m\,\, s}$)} & \colhead{($^{\circ}\,\,^{\prime}\,\,^{\prime \prime}$)} & \colhead{(km$\,$s$^{-1}$)} & \colhead{(Mpc)} & \colhead{(mag)} & \colhead{(\arcsec)} & \colhead{(mag$\,$\arcsec$^{-2}$)} & \colhead{($^{\circ}$)} &  \colhead{(km$\,$s$^{-1}$)} &  \colhead{(km$\,$s$^{-1}$)} & \colhead{(\arcsec)} & \colhead{(\arcsec)} & & \\
 \colhead{(1)} & \colhead{(2)} & \colhead{(3)} & \colhead{(4)} & \colhead{(5)} & \colhead{(6)} & \colhead{(7)} & \colhead{(8)} & \colhead{(9)} & \colhead{(10)} & \colhead{(11)} & \colhead{(12)} & \colhead{(13)} & \colhead{(14)} & \colhead{(15)} 
}
 \startdata
  4257 & 08 10 11.2 & +24 53 32 &   4172 &  59 &  14.0 &  40.3 &  18.7 & 90 & 104 &  86 &   8.9 &   0.05 &  0.22 &  0.00 \\
  4359 & 08 24 33.9 & +74 00 43 &   2272 &  34 &  12.2 &  29.3 &  17.4 & 66 & 125 & 154 &  11.5 &  -0.05 &  0.23 &  0.11 \\
  7687 & 12 32 36.5 & +02 39 37 &   1715 &  34 &  13.0 &  63.0 &  17.5 & 90 &  56 &  53 &  22.2 &   0.04 &  0.12 &  0.05 \\
  8067 & 12 57 12.0 & -01 42 24 &   2833 &  50 &  12.5 &  44.9 &  19.1 & 81 & 126 & 116 &   8.3 &   0.02 &  0.14 &  0.08 \\
  8924 & 14 00 45.6 & +02 01 20 &   3591 &  59 &  13.6 &  39.4 &  19.0 & 90 & 101 & 102 &   8.4 &   0.00 &  0.29 &  0.11 \\
  9187 & 14 21 13.2 & +03 26 08 &   1479 &  28 &  11.0 &  76.9 &  18.4 & 80 & 127 & 117 &  10.3 &   0.01 &  0.12 &  0.05 \\
  9888 & 15 33 05.7 & -01 37 42 &   2805 &  42 &  12.6 &  31.0 &  19.3 & 52 & 127 & 115 &   5.3 &   0.02 &  0.20 &  0.12 \\
 10641 & 16 58 06.4 & +58 53 08 &   5241 &  78 &  14.8 &  32.5 &  19.5 & 90 & 112 & 112 &   9.0 &   0.01 &  0.19 &  0.14 \\
 12641 & 23 30 54.6 & +20 15 02 &   2713 &  39 &  14.0 &  31.4 &  20.0 & 78 &  89 &  68 &   6.2 &   0.06 &  0.30 &  0.07 \\
 20471 & 00 42 14.7 & -18 09 39 &   1536 &  27 &  13.1 &  70.0 &  19.6 & 90 &  85 &  77 &  20.4 &   0.04 &  0.17 &  0.08 \\
 20879 & 01 11 12.5 & -58 50 20 &   4817 &  73 &  14.8 &  30.3 &  20.1 & 90 &  94 &  94 &   8.3 &   0.01 &  0.14 &  0.09 \\
 20893 & 01 12 19.0 & -45 53 17 &   7980 & 115 &  14.0 &  21.0 &  19.3 & 77 & 121 &  85 &   3.0 &   0.06 &  0.37 &  0.22 \\
 20948 & 01 15 52.4 & -54 33 15 &   5450 &  82 &  13.5 &  25.4 &  19.2 & 71 & 118 & 108 &   5.1 &   0.02 &  0.46 &  0.21 \\
 21179 & 01 32 25.7 & -34 36 30 &   5304 &  80 &  13.4 &  20.2 &  18.0 & 67 & 104 & 106 &   8.1 &   0.03 &  0.27 &  0.15 \\
 21214 & 01 32 48.4 & -79 28 25 &   1764 &  29 &  11.6 &  54.1 &  19.5 & 60 &  98 & 114 &  35.7 &   0.07 &  0.08 &  0.03 \\
 21281 & 01 36 33.7 & -80 20 50 &   4107 &  63 &  14.6 &  42.4 &  20.2 & 90 & 103 & 147 &  35.7 &   0.00 &  0.06 &  0.04 \\
\enddata
\tablecomments{The complete version of this table is in the electronic edition of
the Journal.  The printed edition contains only a sample.\\
(1) Uppsala General Catalogue (Nilson 1973) number where available, or else our private Arecibo General Catalogue database number. (2) -- (3) R.A. and Dec., from the literature or measured by us on the POSS-I. (4) Heliocentric (optical) radial velocity measured from best-fit URC curve to long-slit data. See \S~\ref{inversion}. (5) Distance, computed from $V_{\odot}$ and ($\alpha$, $\delta$) using the multi-attractor flow model of Tonry et al. 2000, with $H_o=70\,\,\rm{km\,s^{-1}\,Mpc^{-1}}$. (6) Total apparent I-band magnitude $m_I$, extrapolated to 8 optical scale-lengths (H99). (7) Optical radius $r_{opt}$, corresponding to the radius encompassing 83\% of the total I-band light (H99). (8) Extrapolated I-band disk central surface brightness, corrected for inclination (H99). (9) Inclination obtained from the I-band disk ellipticity (H99) assuming an intrinsic disk axial ratio of 0.2. (10) Rotational velocity, taken as the velocity of the best-fit polyex curve (eq.~(\ref{polyex})) at $r_{opt}$. (11) -- (13) Parameters of best-fit the polyex curve (eq.~(\ref{polyex})) to the RC points. (14) -- (15) Measured value of inner slope $\alpha_m$ and associated error $\Delta \alpha_m$. See \S~\ref{inversion}.
 }
\end{deluxetable*}

 To obtain a sample of galaxies for which the minimum disk assumption is likely valid, we first select by mass, requiring that each system have $V_{rot}<130\,\rm{km\,s}^{-1}$.  This yields 376 dwarf galaxy candidates with high-resolution H$\alpha$ RCs. The adopted upper limit on $V_{rot}$ is arbitrary in the SFI++, but roughly corresponds to the threshold below which disk galaxies lack prominent dust lanes (Dalcanton et al. 2004), have little or no bulges (Kauffmann et al. 2003), and fall systematically below the Tully-Fisher relation (Kannappan et al. 2002). We then examine the optical and I-band emission of each of these candidates, and discard those with distorted morphologies such as strong warps or other irregularities. We also  eliminate systems with evidence for a strong bar or other pathologies deduced from variations in I-band ellipticity, surface brightness, or position angle (PA) across the disk (Haynes et al. 1999, hereafter H99).  We select against the presence of bulges by requiring that the measured total I-band central surface brightness be no more than half a magnitude greater than the extrapolated I-band disk central surface brightness. We do not correct for internal extinction, which for dwarf systems should be quite low (e.g. Giovanelli et al. 1995; Giovanelli \& Haynes 2002). We then examine each galaxy in the Two Micron All Sky Survey Extended Source Catalog and eliminate systems with irregularities or strong bulge contributions not evident in the I-band. These criteria reduce the number of low-mass candidates from 376 to 268. Finally, we discard systems for which the inner RC points do not yield a reliable estimate of the inner halo density because of asymmetries, undersampling, or other distortions, yielding a sample of 165 systems.  

 In principle, one could verify the assumption of a minimum disk by producing multi-component mass models to assess potential contributions of the disk and the bulge to the observed RCs. This is impractical for the present sample, however, because the outer halo shapes (and hence their contribution to the observed RCs) are largely unconstrained by the optical RCs alone. The available I-band photometry constrains the stellar distribution as a function of $r$ across the optical disk. However, the normalization of that distribution may vary by factors of a few given the uncertainties in the stellar mass-to-light ratio (e.g Bell \& de Jong 2001), and for this sample the degeneracy between stellar and halo contributions is exacerbated by the lack of outer halo constraints. Mass models are therefore of little use in determining the baryonic contribution to the optical RC for the systems in our sample, and we do not construct them.

Fig.~\ref{aitoff} shows the sky distribution of the galaxies in the sample, and Table~\ref{alpham} presents their individual properties.
  The majority of the galaxies in our sample (104/165) have long-slit optical spectra from the southern sky survey of Mathewson and collaborators, hereafter MFB (Mathewson et al. 1992; Mathewson \& Ford 1996), who included a larger proportion of dwarf systems in their samples than other contributions to the SFI++. These RCs have also been folded and deprojected by Persic \& Salucci (1995), and thus our sample overlaps with theirs by the same amount (see also Persic et al. 1996). All the galaxies in the sample were observed with a 2\arcsec\ -- wide slit. Fig.~\ref{sam_properties} shows the sample distributions of distances $D$, inclinations $i$, $V_{rot}$ and central I-band surface brightnesses $\mu^I_0$, the latter extrapolated from the disk surface brightness profile and corrected for inclination (H99). The distribution of $\mu^I_0$ in the sample lies in between the LSB and HSB Ursa Major cluster samples of Tully \& Verheijen (1997). A comparison with previous studies deriving $\alpha_m$ via the minimum disk assumption (dB03; S03) is not straightforward, as a conversion from the B- or R-band values listed in these papers is required.  Adopting a canonical conversion of B-I=0.5, however, most of the galaxies of S03 fall within the peak of the $\mu^I_0$ distribution in Fig.~\ref{sam_properties}d. 
The distribution of $\mu^I_0$ thus resembles those in S03 and dB03 even though we did not select based on surface brightness, providing further evidence that the minimum disk approximation is valid for our sample.

There is some disagreement in the literature regarding the range of $i$, $D$ or $M_I$ a galaxy should have to be included in studies of inner halo shapes. 
 For each of these attributes, the parameter space probed by our sample of 165 galaxies is larger than that in previous minimum disk studies: this allows for a systematic investigation of the impact of $i$, $D$ and $M_I$ on the value of $\alpha_m$ returned. Accordingly, we construct a series of subsamples, labeled $k$=1-8, from the 165 galaxies by imposing limits on $i$, $D$, $M_I$ and other parameters. A description of each subsample is given in the first 3 cols. of Table~\ref{KS}.

  Of particular relevance in the measurement of $\alpha_m$ is the distribution of sample galaxy distances (Fig.~\ref{sam_properties}a). Recent CDM simulations predicting the value of $\alpha_{int}$ resolve dark matter halos to $\sim1\%$ of the virial radius, corresponding to the inner 1--2 kpc of low-mass galaxies (Navarro et al. 2003). The reliability of a measured inner halo slope thus depends on the number of resolution elements in the sample RCs on this physical scale. However, the $D$ at which $\alpha_m$ measured via the minimum disk assumption fails to recover the inner halo shape is unclear, since it is muddled by the manner in which raw spectral profiles are smoothed, resampled and folded in the RC derivation (dB01; dB03; S03). The inner regions of the most distant galaxies in this sample ($D\gtrsim150$ Mpc) are poorly resolved, with only 1 or 2 RC points in the inner 2 kpc. For this reason we impose  an upper limit on the range of $D$ in half of the subsamples (3, 4, 7 and 8; see Table~\ref{KS}). Nonetheless, we also analyze the complete sample, which allows for a systematic investigation of the dependence of $\alpha_m$ on $D$ (\S3.1), and provides a consistency check for the simulations we perform (\S4).  The most stringent restriction in $D$ is made for subsample 4 ($D\leq50$ Mpc), which includes 46 systems.  The distribution of $D$ therein most resembles that in dB03 and S03, although the mean $D$ for the samples in these studies is lower than that in subsample 4. There is no overlap between our sample and those in previous studies (dB01; de Blok \& Bosma 2002; Marchesini et al. 2002; dB03; S03; Gentile et al. 2004).

\begin{deluxetable*}{cccc}
\tablecaption{Subsample Descriptions\label{KS}}
\tablewidth{0pt}
\tablehead{
\colhead{Subsample} &  \colhead{Description} & \colhead{\# Galaxies} & \colhead{Med($\alpha_m \pm \Delta \alpha_m$)} \\
\colhead{(1)} & \colhead{(2)} & \colhead{(3)} & \colhead{(4)} \\
}
 \startdata
0 & all data (see \S~\ref{sample}) & 165 & $0.26\,\pm\,0.07$ \\
1 &  $i < 80^{\circ}$ & 91 & $0.27\,\pm\,0.07$ \\
2 & $i < 70^{\circ}$ & 39 &$0.28\,\pm\,0.06$ \\
3 & $D < 100$ Mpc & 124 & $0.23\,\pm\,0.07$ \\
4 &  $D < 50$ Mpc & 46 & $0.22\,\pm\,0.08$\\
5 & $M_I > -20.5$ &  97 & $0.23\,\pm\,0.07$\\
6 & $|x_{off}|<2$\arcsec, $|V^x_{off}|<6 \rm{\,km\,s^{-1}}$ & 91 & $0.27\,\pm\,0.09$ \\
7 & $D < 100$ Mpc, $\delta V < 7\,\rm{km\,s^{-1}}$ & 73 & $0.23\,\pm\,0.07$\\
8 & $i < 85^{\circ}$, $D < 100$ Mpc, $M_I > -21$, $|x_{off}|<3$\arcsec & 46 & $0.25\,\pm\,0.07$ \\
 &   $|V^x_{off}|<10 \rm{\,km\,s^{-1}}$, $\delta V < 8\,\rm{km\,s^{-1}}$ & &  \\
\enddata
\tablecomments{(1) Subsample name. (2) Subsample description. (3) Number of galaxies. (4) 
Median value of  $\alpha_m$, and median error $\Delta \alpha_m$.}
\end{deluxetable*}

\section{Rotation Curve Analysis and Inversion}
\label{inversion}

 As outlined in \S~\ref{sample}, many of the RCs in our sample are mined from previously published studies; the reader is referred to those papers for details on the extraction of rotation velocities from the raw long-slit spectra. Previously unpublished data have been reduced in the same manner as in Catinella (2005). 

Despite some differences in the techniques used to extract RCs from raw long-slit spectra among sample galaxies, the subsequent analysis was performed in a homogeneous manner. 
This is fully described elsewhere (Catinella 2005); here we summarize the procedure. 
We fold each RC about the coordinates $(x_{off},V^x_{off})$, determined via a least-squares fit of the universal rotation curve (URC) of Persic et al. (1996) appropriate for the $M_I$ of the corresponding galaxy. The corrections $(x_{off},V^x_{off})$ are measured relative to the kinematic center $(x_{kin},V_{kin})$ of the RC (Dale et al. 1997), where $V_{kin}$ is the average of the velocities above which 10\% and 90\% of the RC points lie, and $x_{kin}$ is the corresponding position along the slit. This URC folding procedure is superior to one about the kinematic center or about the peak of the continuum emission when either the RC or the galaxy light distribution is asymmetric (Dale et al. 1997, Catinella 2005). 
 The folded RCs are then least-squares fitted with an empirical ``polyex'' curve:
\begin{equation}
V_{pe}(r) = V_o(1-e^{-r/r_{pe}})(1+\beta r/r_{pe})\,\,,
\label{polyex}
\end{equation}
 where $V_o$ sets the amplitude of the fit, $r_{pe}$ is a scale-length that governs the inner RC slope, and $\beta$ determines the outer RC slope. This functional form is sufficiently malleable to model the wide variety of RC shapes found in the SFI++. Our sample galaxies are equally well characterized by a polyex or URC curve; we adopt the former merely for practical purposes, as high-quality polyex fits are archived for all SFI++ galaxies with H$\alpha$ RCs.   The best-fit parameters $V_o$, $r_{pe}$, and $\beta$ for each galaxy in the sample are given in cols. 11--13 in Table~\ref{alpham}. Throughout, we adopt $V_{rot}=V_{pe}(r_{opt})$, where $r_{opt}$ is the radius within which 83\% of the I-band light is contained (H99).

\begin{figure*}
\epsscale{0.95}
\plotone{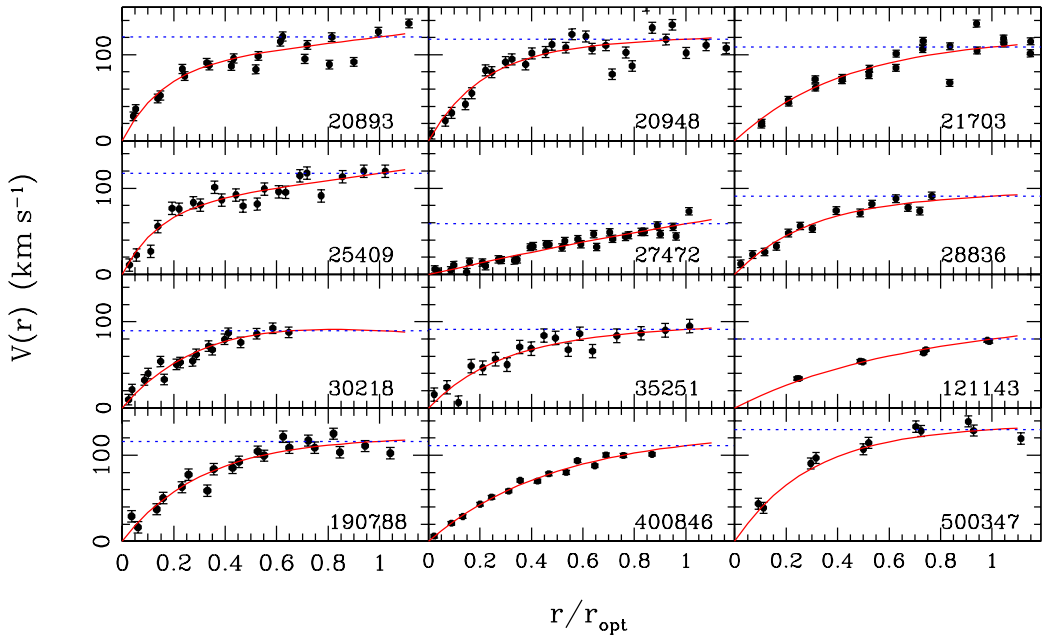}
\caption{Selection of 12 ($i$-corrected) RCs to illustrate the range of sample RC shapes and data quality. The radial coordinate is expressed in units of the optical radius $r_{opt}$, within which 83\% of the I-band light is contained (H99). In each panel, the solid red line is a least-squares polyex fit (eq.~(\ref{polyex})) to the data, the horizontal dotted blue line is the adopted value of $V_{rot}$, and the UGC/AGC number is given in the lower right corner.
 \label{rotcurves}}
\end{figure*}

 As the majority of the RCs in the sample are from MFB, we adjust the properties of the RCs from other sources to match those of the latter. Where required, we drop RC points so that the effective resolution of all the RCs is 2\arcsec\ before folding. As the MFB RCs have no formal error estimates, we adopt the mean deviation of the RC points from the polyex fit within a user-specified $r$ as the error $\delta V(r_i)$ for each RC point. For consistency, we use this definition of $\delta V(r_i)$ even for RCs for which formal error estimates are available. Fig. \ref{rotcurves} shows a selection of 12 RCs from the sample with these uniform properties, chosen to illustrate the range of RC shapes and data quality in the sample.  The best-fit polyex curves for each are over-plotted, $V_{rot}$ is indicated by the horizontal line, and the AGC/UGC number is given in the bottom right corner of each panel.

\begin{figure}
\epsscale{1.25}
\plotone{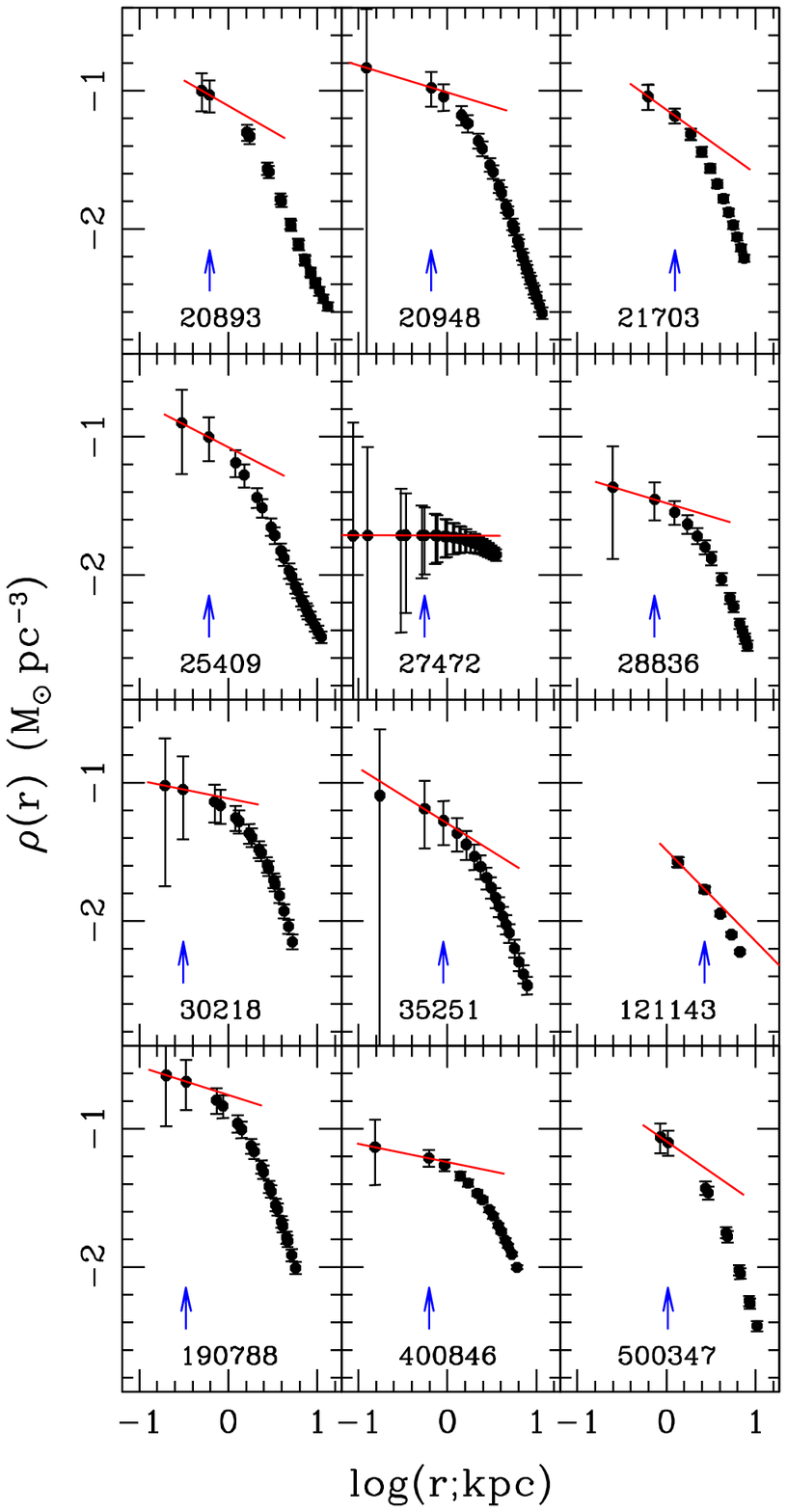}
\caption{Density profiles for the RCs in Fig.~\ref{rotcurves} from eq.~(\ref{polyex}). In each panel, the solid red line shows the linear least-squares fit used to derive $\alpha_m$. The points directly above and left of the blue arrow were included in the fit.  The UGC/AGC number is at the bottom of each panel.
 \label{profiles}}
\end{figure}

Under assumptions of a minimal disk and a spherical halo, the halo density profile can be calculated by inverting the observed RC via a direct application of Poisson's equation:
\begin{equation}
\rho(r) = \frac{1}{4\pi\,G}\left(2\frac{V(r)}{r}\frac{\mathrm{d}V(r)}{\mathrm{d}r} + \frac{V^2(r)}{r^2}\right)\,\,.
\label{invers}
\end{equation}
We note again that it is unphysical to ignore the luminous components completely, even in dwarf and LSB systems (Fall \& Efstathiou 1980). However, in dark-matter dominated systems there is very little change in $\alpha_m$ values when luminous components are included (de Blok \& McGaugh 1997), and minimum disks provide upper limits on the true halo steepness (dB01). In addition, we wish to explicitly test the usefulness of this technique in discriminating between cuspy and corelike halo shapes.

 To compute density profiles using eq.~(\ref{invers}), we assign $V(r_i)$ = $V_{pe}(r_i)$ using the best-fit $V_o$, $r_{pe}$ and $\beta$ in eq.~(\ref{polyex}) at each measured point $r_i$ in the folded RCs. Note that we do not resample the RCs after folding or fitting them, and so the $r_i$ may not be evenly distributed. The derivative of the RC is also estimated from the polyex fit: d$V(r)$/d$r|_{r_i}$ = d$V_{pe}(r)$/d$r|_{r_i}$. We discard points with $r_i < 0.1^{\prime\prime}$ when calculating $\rho(r_i)$, as the fractional errors in $V(r_i)$ for these points are prohibitively large. 

We characterize the inner halo shape of the density profiles as $\rho(r) \sim r^{-\alpha_m}$ for small $r$. For each galaxy, the measured inner slope $\alpha_m$ is estimated by a linear least-squares fit to the inner 2--3 points of a log($\rho (r_i)$) -- log($r_i$) plot, and the errors input into the fit are the larger of the ``up'' and ``down'' errorbars on log($\rho (r_i)$). Because of the large relative errors in log($\rho (r_i)$) for these inner points a least-squares fit is not strictly valid, and the returned formal errors on the fit are meaningless. We therefore estimate an error $\Delta \alpha_m$ in a manner similar to that of dB01, taking $\Delta \alpha_m$ as the mean difference between the reported value and those obtained by including one extra and one fewer point in the fit. Fig.~\ref{profiles} shows the density profiles computed from the RCs in Fig.~\ref{rotcurves}. We derive $\alpha_m$ from the best-fit (solid) line, and points directly above as well as to the left of the arrow are included in the fit. The UGC/AGC number for each galaxy is at the bottom of each panel. Fig.~\ref{profiles} shows that in general $r_{br} < 1\,$kpc, where $r_{br}$ is the outermost point included in the fit.

\subsection{Results for the Galaxy Sample}
\label{res_sample}

\begin{figure}
\epsscale{0.9}
\plotone{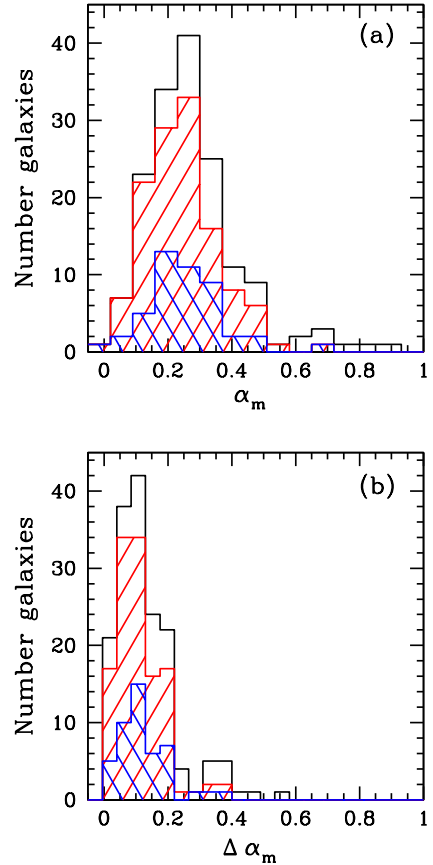}
\caption{Distributions of a) $\alpha_m$ and b) $\Delta \alpha_m$ for all sample galaxies (empty histogram), galaxies in subsample 3 (hatched red histogram) and galaxies in subsample 8 (cross-hatched blue histogram). The median values of $\alpha_m$ and $\Delta \alpha_m$ for all subsamples are given in Table~\ref{KS}.
 \label{singlehist}}
\end{figure}

 The values of $\alpha_m$ and $\Delta \alpha_m$ obtained for each galaxy in the sample are given in cols. 14 and 15 of Table~\ref{alpham}, and histograms of these quantities for subsamples 0, 3 and 8 are shown in Fig.~\ref{singlehist} and illustrate the distributions of $\alpha_m$ for a variety of selection criteria. The histograms corresponding to the other subsamples are similar in shape, and the median values of $\alpha_m$ and $\Delta \alpha_m$ for each are given in col. 4 of Table~\ref{KS}.
When all 165 galaxies in the sample are included, the median values of the measured inner slope and corresponding uncertainty are $\alpha_m = 0.26 \pm 0.07$. The median values for the 8 subsamples are similar, and in agreement within the median $\Delta \alpha_m$. The subsample with the lowest median, $\alpha_m = 0.22 \pm 0.08$ (subsample 4), is that containing only the nearest galaxies: this is in good agreement with the results of dB01 and dB03, who find a distribution peaked at $\alpha_m=0.2$ with a tail toward steeper slopes. 

 For our sample, Fig.~5 illustrates that some of the asymmetry towards steeper slopes in the distribution of $\alpha_m$ for subsample 0 stems from the inclusion of systems with $D>100$~Mpc, as the ``end'' of the tail ($\alpha_m \gtrsim 0.6$) disappears in the distribution for subsample 3. There is no clear change in histogram morphology when further parameter restrictions are imposed (e.g. subsample 8).  For all subsamples, $\Delta \alpha_m$ is typically a significant fraction of $\alpha_m$; each individual estimate of $\alpha_m$ thus has little significance.  The distribution of values, on the other hand, does place some constraints on the range of inner halo shapes measured. It is clear from Fig.~\ref{singlehist}a and Table~\ref{KS} that the measured inner slopes $\alpha_m$ are much shallower than the intrinsically cuspy halos ($\alpha_{int} \sim 1$) predicted by CDM simulations of structure formation. The data thus exhibit the cusp/core problem.

\begin{figure}
\epsscale{1.2}
\plotone{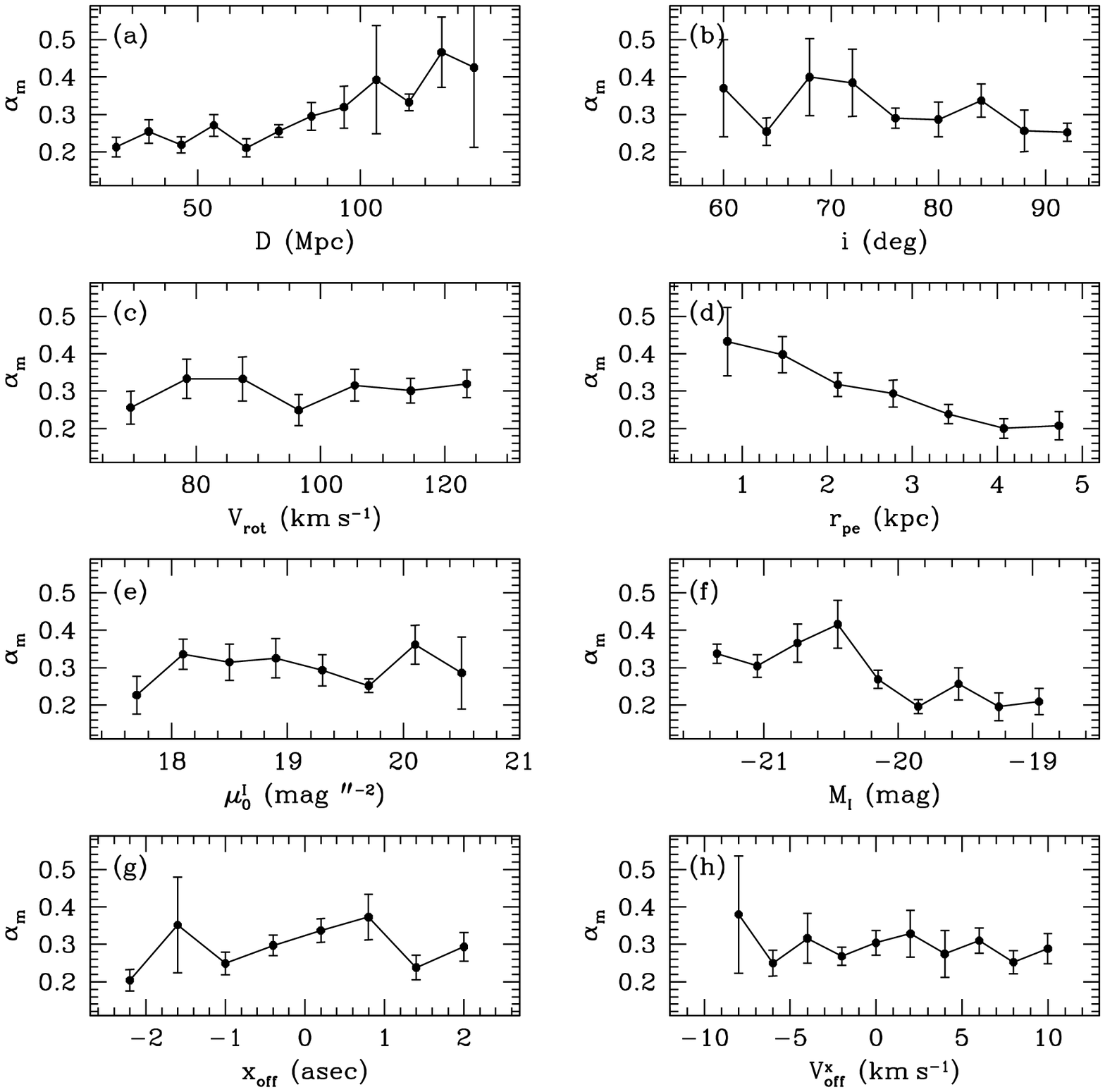}
\caption{Mean value of $\alpha_m$ for the sample galaxies as a function of a) $D$, b) $i$, c) $V_{rot}$, d) $r_{pe}$, e) $\mu^I_0$, f) $M_I$, g) $x_{off}$, and h) $V^x_{off}$. The points show bin centers, and errorbars are given by  $\sigma_j/\sqrt{n_j}$, where $\sigma_j$ is the standard deviation of the mean for $n_j$ values of $\alpha_m$ in the $j^{th}$ bin.
\label{sam_trends}}
\end{figure}

 In Fig.~\ref{sam_trends} we show mean values of $\alpha_m$ for the sample as a function of other galaxy parameters. The points show the bin centers with errorbars given by  $\sigma_j/\sqrt{n_j}$, where $\sigma_j$ is the standard deviation of the mean for $n_j$ values of $\alpha_m$ in the $j^{th}$ bin. There is a clear trend of increasing $\alpha_m$ with increasing $D$, which we explore in more detail below. 
There is also a decrease in $\alpha_m$ with increasing $r_{pe}$: the steeper the inner RC rise as characterized by the polyex fit, the larger the measured value of $\alpha_m$ (see \S~\ref{bias}). 
In addition, there is some evidence that galaxies at higher $i$ have lower $\alpha_m$, as found by Rhee et al. (2004). This could be the result of greater slit smearing at higher inclinations, extinction effects or baryonic distortions hidden in edge-on disks and hence not apparent in the I-band photometry. There is also some indication that galaxies with $M_I < -20.5$ (as in subsample 5) have larger $\alpha_m$ than fainter systems, which may signal a non negligible contribution from the stellar disk in the highest mass galaxies in the sample. However, no corresponding trend is found between $\alpha_m$ and $V_{rot}$, nor between $\alpha_m$ and $\mu^I_0$.  Similarly, there is no obvious dependence of $\alpha_m$ on the offsets $x_{off}$ or $V^x_{off}$ obtained during the data processing stage (see \S~\ref{inversion}). 

\begin{figure}
\epsscale{1.2}
\plotone{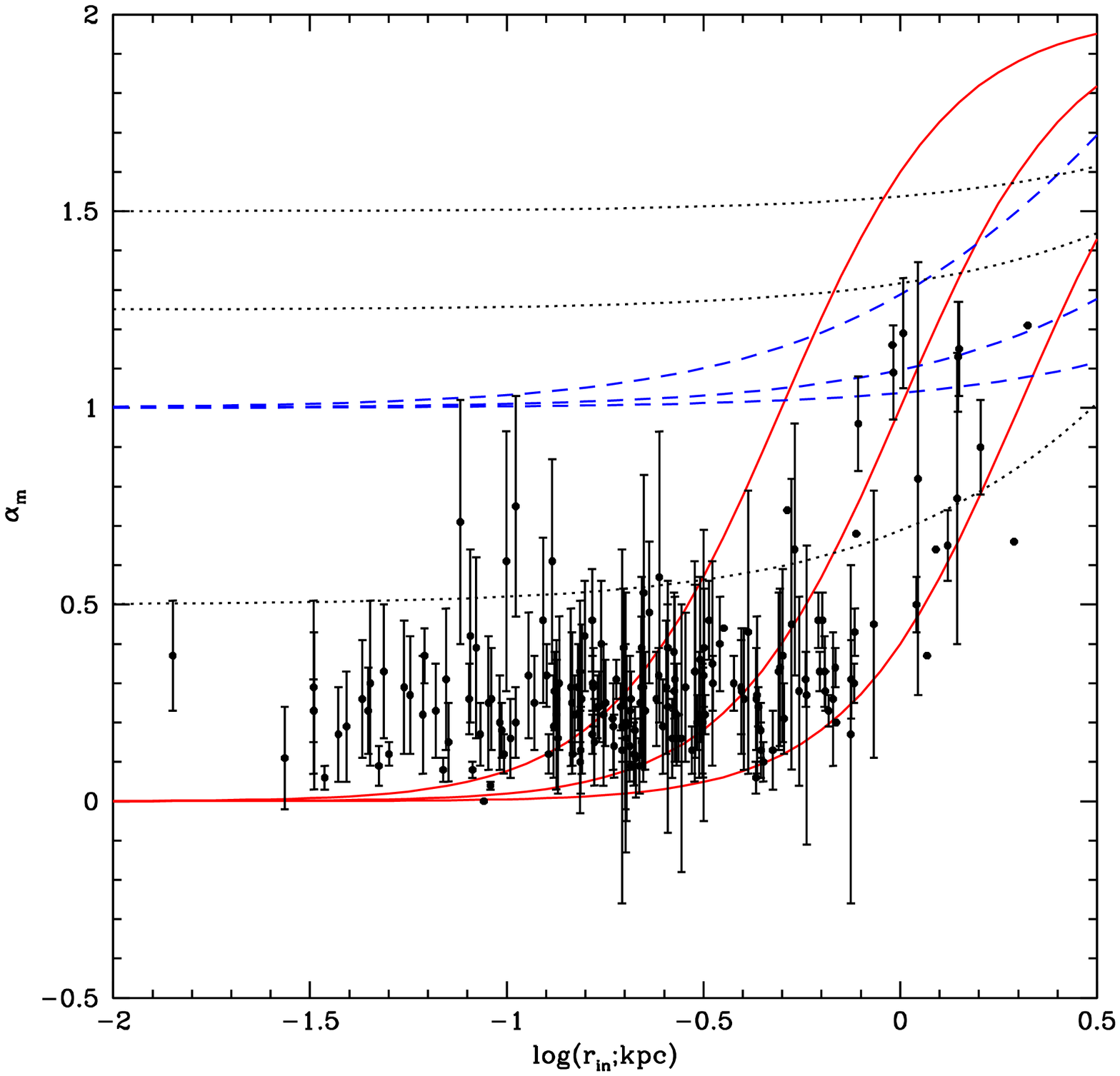}
\caption{Inner slopes $\alpha_m \pm \Delta \alpha_m$ for the sample galaxies as a function of the innermost point $r_{in}$ of the folded RCs. Points for which errorbars are not shown have $\Delta \alpha_m$ smaller than the point size (see Fig.~\ref{singlehist} and Table~\ref{alpham}).
Solid red lines show $|\mathrm{d} \log \rho/ \mathrm{d} \log r|$ evaluated at $r_{in}$ for a pseudo-isothermal halo with $r_c=0.5$ (left curve), $r_c=1$ (middle curve), and $r_c=2$ (right curve; see \S~\ref{res_sample})). Dashed blue lines show expectations for NFW halos (eq.~(\ref{NFW})) consistent with the range of $V_{rot}$ in the sample (see \S~\ref{alpha1}): $c=12,\,V_{200}=50\,\rm{km\,s}^{-1}$ (top curve), $c=8,\,V_{200}=110\,\rm{km\,s}^{-1}$ (middle curve) and $c=5,\,V_{200}=180\,\rm{km\,s}^{-1}$ (bottom curve). Dotted lines show $|\mathrm{d} \log \rho/ \mathrm{d} \log r|$ for halos with similar to NFW (eq.~(\ref{density})), but with $\alpha_{int} = 1/2$ (bottom), $\alpha_{int} = 5/4$ (middle) and  $\alpha_{int} = 3/2$ (top; see \S~\ref{alphane1}).
\label{rin}}
\end{figure}

 The effect of RC resolution on $\alpha_m$ is illustrated in Fig.~\ref{rin}, where $\alpha_m \pm \Delta \alpha_m$ is plotted as a function of $r_{in}$, the location of the RC point that lies closest to $r=0$ expressed in kpc. The solid lines show the expected slope $\alpha =|\mathrm{d}\log \rho / \mathrm{d}\log r|$ (evaluated at $r_{in}$) for an isothermal model $\rho \propto [1 + (r/r_c)^2]^{-1}$ often used to parameterize halo shapes. From left to right, models with $r_c=$~0.5, 1 and 2 are overplotted. The dashed lines show $\alpha$ expected for NFW halo shapes (eq.~(\ref{NFW})) that bracket the range of $V_{rot}$ in the sample (see \S~\ref{alpha1}): from top to bottom, models with ($c=12,\,V_{200}=50\,\rm{km\,s}^{-1}$), ($c=8,\,V_{200}=110\,\rm{km\,s}^{-1}$), and ($c=5,\,V_{200}=180\,\rm{km\,s}^{-1}$) are overplotted. The dotted lines show CDM halo shapes (eq.~(\ref{density})) analogous to the middle dashed line, but with $\alpha_{int}=1/2$ (bottom), $\alpha_{int}=5/4$ (middle), and $\alpha_{int}=3/2$ (top; see \S~\ref{alphane1}).  Qualitatively, the distribution of points in Fig.~\ref{rin} is very similar to that in fig.~3 of dB01 (see also fig. 14 in de Blok \& Bosma 2002 and fig. 12 of S03), despite significant differences in the manner in which $r_{in}$ is derived. In particular, the RCs of dB01 are smoothed and re-sampled after folding whereas the ones in the present study are not.

 From Fig.~\ref{rin} it is clear that the $\alpha_m$ values in our sample are in better agreement with the expected slopes of pseudo-isothermal halos than any of the NFW-type halos, and follow the expected trend of increasing $\alpha_m$ in the pseudo-isothermal models for log($r_{in}$)$>-0.3$.  At small $r_{in}$, however, the measured values do not converge to $\alpha_m \sim 0$, as found by dB01, but remain relatively constant for $-1.5 \lesssim \log (r_{in}) \lesssim -0.3$: a characteristic $\alpha_m \sim 0.25$ is measured throughout this range.
 The dependence of $\alpha_m$ on $r_{in}$ is clearly a resolution effect, first noted by dB01. More distant systems are more poorly sampled than those of nearby galaxies in our long-slit spectra, resulting in the measurement of $\alpha_m$ over a larger extent of the disk. A correlation between $\alpha_m$ and $D$ is therefore expected from the shape of $\rho (r_i)$ (Fig.~\ref{profiles}), since the slope of the density distribution steepens as $r$ increases. This same phenomenon is likely responsible for the lower median $\alpha_m \pm \Delta \alpha_m$ in subsamples 3, 4 and 7, the tail toward larger $\alpha_m$ for subsample 0 in Fig.~\ref{singlehist}a, and the trend between $D$ and $\alpha_m$ in Fig.~\ref{sam_trends}a. The need for spatial resolution in determining the halo shapes of galaxies is clear in Fig.~\ref{rin} and underscores the importance of taking resolution effects into account when interpreting the results. The small, relatively constant value of $\alpha_m$ obtained for log($r_{in}$)$<-0.3$ also agrees with Salucci's (2001) conclusion that all spiral galaxies have large core radii.

 The discrepancy between the $\alpha_m$ obtained and the cusps predicted by the CDM paradigm is actually greater than Figs.~\ref{singlehist} and \ref{rin} imply for a number of reasons. Recall that the $\alpha_m$ were derived for minimum disks, an assumption that yields an upper limit on the true halo steepness. In addition, $|\mathrm{d}\log \rho / \mathrm{d}\log r|$ values plotted for the NFW halos in Fig.~\ref{rin} do not take the impact of the baryon collapse on the halo shape into account, which tends to increase $\alpha_{int}$ over the values determined from collisionless simulations (Blumenthal et al. 1986; Flores et al. 1993; Gnedin et al. 2004). Recent simulations also indicate that $\alpha_{int}$ from collisionless simulations is larger than the NFW value for low-mass galaxies (Navarro et al. 2003); the dotted line corresponding to $\alpha_{int}=5/4$ in Fig~\ref{rin} may therefore better reflect CDM predictions than NFW halos. Finally, $\alpha_m$ is measured from a fit to points distributed between $r_{in}$ and $r_{br} \sim 1\,$kpc, and Figs.~\ref{profiles} and \ref{rin} show that $|\mathrm{d}\log \rho / \mathrm{d}\log r| \ge \alpha_{int}$ for $r>0$. However, there is little change in the middle-dashed and dotted lines of Fig.~\ref{rin} for $r \lesssim 1\,$kpc, where $\alpha_m$ is typically measured (Fig.~\ref{profiles}). Comparing $\alpha_m$ directly to $\alpha_{int}$ is therefore a reasonable approximation for the majority of the sample galaxies, and adequate for the qualitative analysis performed here.

\section{Simulations}
\label{simulations}

The results of \S\ref{res_sample} demonstrate that in general, the galaxies in our sample have $\alpha_m$ values that are much more core-like than the cuspy $\alpha_{int}$ values that CDM predicts. 
The data are clearly consistent with a scenario in which low-mass galaxies have intrinsic cores rather than cusps.
However, a variety of observational effects systematically lower $\alpha_m$ relative to $\alpha_{int}$ to varying degrees if the latter is non-zero.
Conversely, the dependence of $\alpha_m$ on $D$ evident in Figs.~\ref{singlehist}, \ref{sam_trends}, and \ref{rin} demonstrates that the properties of the sample itself may also influence the values of $\alpha_m$ obtained. We also do not account for $|\mathrm{d}\log \rho / \mathrm{d}\log r| \ge \alpha_{int}$ at the location where $\alpha_m$ is measured in the comparisons of \S~\ref{res_sample} (Fig.~\ref{rin}). It is therefore necessary to assess the extent of the potential biases and resolution effects on $\alpha_m$ in order to understand the implications of our results for the CDM paradigm.

 In this section we investigate the nature of the cusp/core problem by simulating observations of model galaxies with various $\alpha_{int}$ and by comparing outputs to the distributions of $\alpha_m$ obtained in each of the subsamples. These simulations have a single goal: to determine whether long-slit observations, in conditions typical of those in the SFI++ and of model galaxies with cuspy $\alpha_{int}$ that are broadly consistent with the predictions of the CDM paradigm, may yield a distribution of $\alpha_m$ similar to that obtained for the data. We require that the model galaxies have attributes similar to those of each subsample (e.g. Fig.~\ref{sam_properties}), and therefore simulations that recover the distribution of $\alpha_m$ should also recover trends with other measurable properties (Fig.~\ref{sam_trends}). The NFW profile has been extensively discussed in the literature (see \S~\ref{intro}). We therefore concentrate on the $\alpha_{int}=1$ case, both because of its relevance to other work and because prescriptions for generating NFW halos appropriate for low-mass galaxies are readily available. We then modify our technique to consider models with $\alpha_{int}=1/2,\,5/4$ and 3/2.

 We review the potential biases on $\alpha_m$ that we examine in our simulations in \S~\ref{bias}. In \S~\ref{model} and \S~\ref{obs}, we describe the model galaxies and mock long-slit observations used in the simulations, respectively.

\subsection{Potential Biases on $\alpha_m$}
\label{bias}
The uncertainties inherent in obtaining long-slit spectra along the major axes of low-mass galaxies, as well as the galaxy geometries themselves, may bias the shape of the resulting RCs.  The impact that these biases then have on $\alpha_m$ is illustrated by considering the relationship between $\alpha_m$ and the RC scalelength $r_{pe}$ of the polyex function for our sample. From the values of $\beta$ in Table~\ref{alpham} and the form of eq. (\ref{polyex}), for $r \lesssim r_{pe}$ we can approximate:
\begin{equation}
\frac{V_{pe}(r)}{V_0} \simeq 1 - e^{-r/r_{pe}} = 1-P(r)\,\,\,.
\label{lim}
\end{equation} 
 Substituting this into equation~(\ref{invers}) and computing $\alpha=|\mathrm{d}\log \rho / \mathrm{d}\log r|$ gives:
\begin{equation}
\alpha= \frac{2}{r_{pe}} \left( \frac{r^2[P(r)-2P(2r)] + r_{pe}^2[1-2P(r)+P(2r)]}{(P(r)-1)[x(P(r)-1) - 2rP(r)]}\right)\,\,\,.
\label{rpe_alpha}
\end{equation}
 In equation~(\ref{rpe_alpha}), the second fraction is positive and varies little with $r_{pe}$ for $r \lesssim 1\,$kpc. We therefore expect $\alpha_m$ to decrease with increasing $r_{pe}$ for the sample galaxies, as is clearly the case in Fig.~\ref{sam_trends}d. An analogous effect will occur for $\alpha_m$ measured for a single galaxy, if $r_{pe}$ is systematically increased. Eq. (\ref{lim}), the shape of the RCs in Fig.~\ref{rotcurves} and their corresponding polyex fits (Table~\ref{alpham}) show that a decrease in the amplitude and/or derivative of $V(r)$ in the inner RC regions results in larger $r_{pe}$. Qualitatively, then, a mechanism that biases $V(r)$ or its derivative low relative to its true value will in turn lead to an underestimate of $\alpha_m$. 

 A number of potential RC biases have been discussed by S03 and dB03. Here, we review those addressed by the simulations described in \S~\ref{model} and \S~\ref{obs}. First, the finite angular resolution of the observations as well as the finite width of the slit lead to the inclusion of gas away from the major axis in the long-slit spectra. The projected radial velocities away from the major axis are lower than those along it; the gas at these locations may therefore lower the $V(r)$ fit to the emission line shape along the slit. The same is true if the slit is not aligned with the galaxy major axis, with the impact on the RC shape exacerbated by the lower RC derivative away from the major axis. In addition, the RCs of highly inclined systems may have low $V(r)$ and derivative because the disk geometry forces the inclusion of more gas at low projected velocities along the line of sight of the slit than for systems at lower $i$. Since all these effects bias $V(r)$ and its derivative low these uncertainties will add coherently in real systems, and from eq.~(\ref{rpe_alpha}) we anticipate that $\alpha_m < \alpha_{int}$. 

 The magnitudes of these effects have been analyzed by S03 (see their fig.~8), who measure $\alpha_m$ for model galaxies with a range of distances and inclinations, and ``observed'' with a variety of slit offsets. For halos with $\alpha_{int} \sim 1$, $\alpha_m$ depends most strongly on the degree of slit mis-alignment.  Given that slit offsets are generally not directly measurable (see \S~\ref{obs}), that each estimate of $\alpha_m$ is highly uncertain (Fig.~\ref{singlehist}), and that the models are necessarily simple in comparison to real systems (see \S~\ref{model}), these simulations cannot be used to reliably correct $\alpha_m$ on a case-by-case basis (S03; dB03). Nonetheless, the impact of these biases can be assessed from a statistical standpoint by simulating observations in conditions similar to those in which the sample was derived, and using model galaxies with properties that reflect those of the sample. 

 We note that the resolution effects discussed in \S~\ref{res_sample} will bias $\alpha_m$ in the opposite sense from that discussed above. To account for this, the distribution of $D$ for the model galaxies is the same as that for the subsamples, and the same algorithms are used to derive $\alpha_m$ for both the model and actual RCs. Our simulation outputs therefore include resolution effects in addition to potential observing and geometric biases.

 There are also more complex factors that bias RCs derived from long-slit spectra but are beyond the scope of our simulations: examples are irregular H$\alpha$ distributions, non-circular motions, finite disk thicknesses and small, undetected bulges or bars in the sample systems. These last three effects are examined in detail by Rhee et al. (2004), who use N-body simulations of dark and stellar components to assess the difference between the measured $V(r)$ and the actual circular velocity along the major axis. For most disk geometries, these factors also lower the value of $\alpha_m$ derived relative to $\alpha_{int}=1$. Since these effects are likely present in our sample but not in the models, our simulation outputs represent conservative estimates of the biases inherent in measuring $\alpha_m$.

\subsection{Galaxy Models}
\label{model}
We model dwarf galaxies by embedding infinitely thin, uniform H$\alpha$ disks in spherically symmetric halos with density distributions given by:
\begin{equation}
\rho(r) = \frac{\rho_o}{(cx)^{\alpha_{int}}(1+cx)^{3-\alpha_{int}}}\,\,,
\label{density}
\end{equation}
where $x = r/r_{200}$, $c$ is the concentration index of the halo, $\rho_o$ is a characteristic halo density and $r_{200}$ is the halo size, corresponding to the radius within which $\overline{\rho(r)} = 200\rho_{crit}$, with $\rho_{crit}$ the critical density for closure of the universe. For this family of density distributions, the corresponding halo circular velocity curve is:
\begin{equation}
V_h(r) = V_{200}\sqrt{\frac{\mu(cx)}{x \mu(c)}}\,\,,
\label{vc}
\end{equation}
with 
\begin{equation}
\mu(z)=\int_0^zy^{2-\alpha_{int}}(1+y)^{\alpha_{int}-3}dy
\label{mu}
\end{equation}
(Zhao 1996) and, by definition, $V_{200}=h r_{200}$. In each simulation, we generate a set of halos with a variety of $V_{200}$ and $c$ but with the same $\alpha_{int}$. We elaborate on how $V_{200}$ and $c$ are chosen for model galaxies with $\alpha_{int} = 1$ in \S~\ref{alpha1} and with $\alpha_{int} \ne 1$ in \S~\ref{alphane1}.

Once the halo is specified, primary galaxy parameters such as $D$ and $i$ are chosen as random deviates of the corresponding distributions for the subsample to which the simulations will be compared. The I-band magnitude $M_I$ is not explicitly required in the derivation of $\alpha_m$ but provides an {\em ansatz} for the least-squares fitting routines in the RC folding process described in \S~\ref{inversion}, and as such is assigned to each model in the same manner.  
We assume that the presence of the disk has no impact on the halo shape, and do not apply adiabatic contraction corrections to our models.  

\subsubsection{Halos with $\alpha_{int}=1$: NFW Profiles}
\label{alpha1}
For $\alpha_{int}=1$ in eq.~(\ref{density}) we recover the NFW halo profile in eq.~(\ref{NFW}). In this case the halo velocity curve in eq.~(\ref{vc}) has an analytic solution:
\begin{equation}
\left[\frac{V_h^{NFW}(r)}{V_{200}}\right] ^2 = \frac{1}{x} \frac{\ln({1+cx}) -(cx)/(1+cx)}{\ln({1+c}) - c/(1+c)}\,\,.
\label{Vnfw}
\end{equation} 
Given the minimum disk analysis of \S~\ref{inversion}, we assume that the halo is the sole contributor to the kinematics of the model galaxies. For each halo the input $V_{200}$ is therefore determined by substituting $V_h^{NFW}(r_{opt})=V_{rot}$ into eq.~(\ref{Vnfw}) and solving for $V_{200}$\footnote{Mo et al. (1998) and Seljak (2002) determine $V_{200}-V_{rot}$ relations for $L_*$ galaxies by examining the observed Tully-Fisher relation and its scatter. The galaxies in our sample generally have much smaller masses than these systems, however, as well as substantially submaximal disks, and as such these relations are not applicable here.}, given a relationship between $V_{200}$ and $c$.

Observationally, the characteristic halo concentration derived from fitting mass models to hybrid $H\alpha$+\ion{H}{1} RCs of dwarf and LSB galaxies is $c \sim 6$ (McGaugh et al. 2003). Since we wish to probe the relationship between the observed $\alpha_m$ and theoretical CDM halo shapes, however, we adopt a distribution of $c$ that is consistent with predictions from N-body simulations. 
 NFW show that the value of $c$ for a halo depends on the mean density of the universe at the time of halo formation, and provide a prescription for computing $\log c$ given the halo mass and a specific cosmology. We perform this computation for a cosmology consistent with recent {\it Wilkinson Microwave Anisotropy Porbe} results (Spergel et al. 2003): $\Omega_m=0.27$, $h=0.7$ and $\sigma_8=0.9$. Jing (2000) shows that the distribution in $c$ can be well described by a lognormal function for a given halo mass, and Wechsler et al. (2002) find a standard deviation of $\sigma_{\log c} = 0.12$ for present-day halos with well-constrained $c$. 

\begin{figure}
\epsscale{1.2}
\plotone{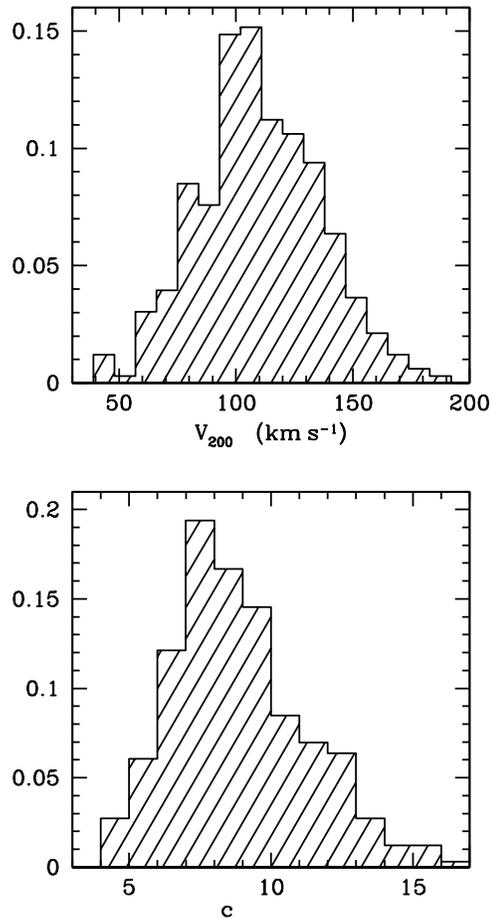}
\caption{Typical distributions of $V_{200}$ (top) and $c$ (bottom) for model galaxies in simulations; see \S~\ref{alpha1} for details.
\label{inputs}}
\end{figure}

 We therefore determine $V_{200}$ and $c$ for each model galaxy as follows. First, a ($V_{rot}, r_{opt}$) pair measured for one of the galaxies in the subsample of interest is randomly chosen.  We then solve eq.~(\ref{Vnfw}) for $V_{200}$, by substituting $V_h^{NFW}(r_{opt})=V_{rot}$ and using the $V_{200} - \log c$ relation from NFW. The value of $c$ adopted for the halo is that from the relation, with a scatter in $\log c$ that is a Gaussian random deviate with $\sigma_{\log c} = 0.12$. Typical distributions of input $V_{200}$ and $c$ determined in this manner are shown in Fig.~\ref{inputs}, and the dashed lines in Fig.~\ref{rin} show $|\mathrm{d} \log \rho/ \mathrm{d} \log r|$ for the corresponding range of halo shapes. The distributions of $V_{rot}$ returned by the simulations resemble those in the data, suggesting that our method for selecting $V_{200}$ does indeed yield appropriate halo shapes. In addition, changes in $V_{200}$ by $\lesssim$50 $\rm{km\,s^{-1}}$ affect primarily the amplitude of the observed RC (eq.~\ref{Vnfw}), by an amount less than 25 $\rm{km\,s^{-1}}$ at $r_{opt}$; since the inner halo shape should depend little on the RC amplitude, we expect that variations in the method adopted to determine $V_{200}$ will not significantly alter our results (see \S~\ref{int1}).

\subsubsection{Halos with $\alpha_{int} \ne 1$}
\label{alphane1}

To examine the impact of our observing and data processing techniques on a range of cuspy intrinsic halo shapes, we generate halos with inner slopes that are shallower ($\alpha_{int}=1/2$) and steeper ($\alpha_{int}=5/4$ and 3/2) than the NFW value. For $\alpha_{int}=1/2$ and 3/2, eq.~(\ref{mu}) has an analytic solution, and the halo velocity curves are given by
\begin{equation}
\left[\frac{V_h^{1/2}(r)}{V_{200}}\right] ^2 = \frac{1}{x} \frac{\ln({1+2cx+2S(cx)}) -2R(cx)/T(cx)}{\ln({1+2c+2S(c)}) - 2R(c)/T(c)}
\label{V1/2}
\end{equation}

\begin{equation}
\left[\frac{V_h^{3/2}(r)}{V_{200}}\right] ^2 = \frac{1}{x} \frac{\ln({1+2cx+2S(cx)}) -(2cx)/S(cx)}{\ln({1+2c+2S(c)}) - (2c)/S(c)}\,\,,
\label{V3/2}
\end{equation}
respectively, where $R(x)= x(1+4/3x)$, $S(x)=\sqrt{x(1+x)}$ and $T(x)=(1+x)S(x)$. We also simulate galaxies with $\alpha_{int}=5/4$, a value that is a better approximation to the inner halo slope near $r_i \sim 1\,\rm{kpc}$ than the NFW value for low-mass halos (Navarro et al. 2003; see \S~\ref{intro}). For $\alpha_{int}=5/4$, we solve eq.~(\ref{mu}) by quadrature.

 There are no prescriptions available in the literature for determining $c$ from $V_{200}$ for galaxy halos when $\alpha_{int}\ne1$, so we cannot use eqs. (\ref{V1/2}) and (\ref{V3/2}) directly. We therefore determine these parameters by ``bootstrapping'' from the $\alpha_{int}=1$ case under the assumption that the family of halos in eq.~(\ref{density}) is self-similar, such that the peak of the circular velocity profile ($V_{max}$) occurs at the same value of $r$ ($r_{max}$) for halos with different $\alpha_{int}$. For each randomly drawn $V_{rot}$, we assign an NFW halo shape according to the prescription in \S~\ref{alpha1}; this specifies an $r_{max}$ and $V_{max}$. We then choose $V_{200}$ and $c$ for the desired $V^{\alpha_{int}}_h(r)$ by requiring that its $r_{max}$ and $V_{max}$ match the specified NFW profile. Then, we scale all the $V^{\alpha_{int}}_h(r)$ values in a simulation by the same small factor ($\le 20$\%), so that the centroid of the distribution of $V_{rot}$ measured for the simulated RCs is comparable to that for each subsample.  For a range of $\alpha_{int}$ near 1 ($1/2\,\lesssim\,\alpha_{int}\,\lesssim\,3/2$) this method generates halos with a variety of shapes that are broadly consistent with the current CDM paradigm and that have the same statistical properties as the data. The dotted lines in Fig.~\ref{rin} show $|\mathrm{d} \log \rho/ \mathrm{d} \log r|$ for halo shapes consistent with the distribution of $V_{rot}$ in the subsamples.

 We note that the RC shapes for self-similar halos with $\alpha_{int} \lesssim 0.5$ or $\alpha_{int} \gtrsim 1.5$ are quite different from those with $\alpha_{int}=1$, and the scaling of $V^{\alpha_{int}}_h(r)$ required to match the distribution of $V_{rot}$ in the sample becomes prohibitively large. We therefore do not generate halos with $\alpha_{int}=0$, since the goal of the simulations is to compare our data with ``observed'' halos consistent with the CDM paradigm (which $\alpha_{int}=0$ halos are not), and because such halos would have to be generated via a different algorithm from the one presented here.

\subsection{Long-Slit ``Observations'' and Analysis}
\label{obs}

 We obtain RCs for each of the model galaxies by ``observing'' them with a long slit. We adopt slit dimensions similar to those used to obtain long-slit spectra for the sample galaxies:
the slit is 120\arcsec\ long and 2\arcsec\ wide, with 0.66\arcsec\ pixels. The instrumental response is taken to be Gaussian, with a FWHM of $30\,\rm{km\,s^{-1}}$. We assume a seeing FWHM of 1\arcsec\ throughout.
 To derive the spectral profile for a model galaxy we center the mock slit on the inclined disk, but with some error in the slit position on the sky ($\mathbf{r_{off}}$) relative to the galaxy center, error in the PA relative to the galaxy major axis ($\Delta PA$), and error in the spectral center of the slit relative to the galaxy's recessional velocity ($V_{off}$).

 In order for our simulation to reproduce any biases in $\alpha_m$ that may be present in our data, the values of $\mathbf{r_{off}}$, $\Delta PA$ and $V_{off}$ should reflect the observing conditions under which the sample RCs were obtained. However, direct measurements of these quantities are not available. This is particularly the case for $\mathbf{r_{off}}$, as it depends on {\it (i)} the pointing accuracy of the telescope, and {\it (ii)} the accuracy with which the true RC center can be placed in the slit. The former has a well-determined value, but the latter will depend on a variety of factors such as the characteristics of the galaxy light and dust distribution, the band in which the photometric center of the galaxy is estimated prior to the observations, the disk orientation and the seeing. Factor {\it (ii)} is of particular importance for our sample, as the selection criteria in \S~\ref{sample} generally favor galaxies with photometric centers that are more poorly estimated than those of $L_*$ spiral galaxies.

\begin{figure}
\epsscale{1.2}
\plotone{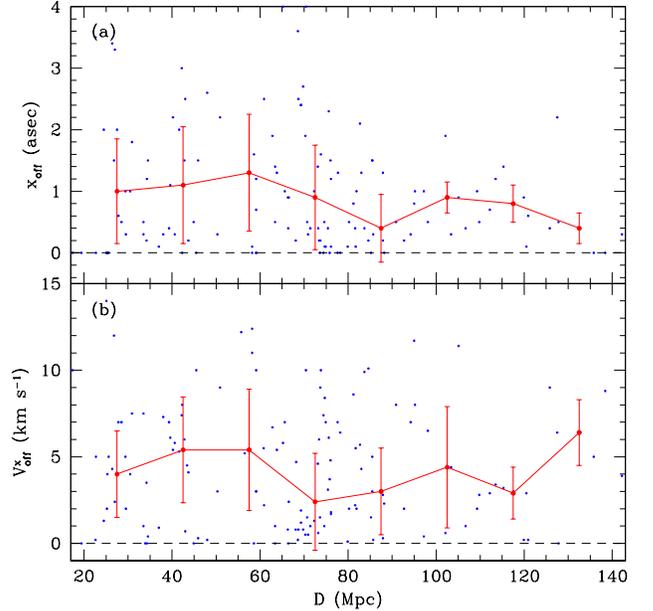}
\caption{Variation of a) $|x_{off}|$ and b) $|V^x_{off}|$ with $D$ for all sample galaxies. Small blue dots show individual datapoints, and large, connected red circles show the median value at the corresponding bin center. Errorbars on the large points show the interquartile range. By definition, all points must lie above the dashed line.
\label{xoffdist}}
\end{figure}

\begin{figure}
\epsscale{1.2}
\plotone{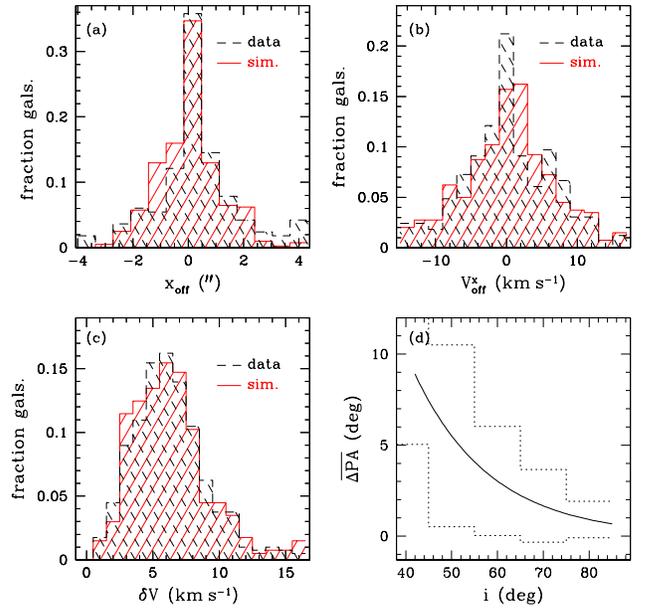}
\caption{Distributions of a) $x_{off}$, b) $V^x_{off}$, and c) $\delta V$ for sample RCs (dashed lines) and the corresponding simulated RCs (solid red lines). d) $\overline{\Delta PA(i)}$ (solid line) with uncertainty $\sigma_{PA}(i)$ (dotted line) as a function of $i$. In the simulation, $\Delta PA(i)$ is a Gaussian random deviate with mean $\overline{\Delta PA(i)}$ and standard deviation $\sigma_{PA}(i)$.
 \label{errors}}
\end{figure}

 To estimate $\mathbf{r_{off}}$, we use the distribution of offsets $x_{off}$ and $V^x_{off}$ for the galaxy sample, determined via a best-fit URC curve during the data processing stage. As outlined in \S~\ref{inversion},  $x_{off}$ and $V^x_{off}$ are measured relative to the galaxy kinematic center, which itself results from the average of the velocities above which 90\% and 10\% of the RC points lie. The parameters $x_{off}$ and $V^x_{off}$ are thus measures of RC asymmetry. In real systems this asymmetry may stem from irregularities in the RC itself (due to a strong bar, patchy H$\alpha$ emission, or differing RC extents on either side of the nucleus, for example), as well as from a misplacement of the true RC center and galaxy major axis relative to the slit. Since we select against the presence of RC distortions in the construction of the dwarf galaxy sample, their impact on $x_{off}$ and $V^x_{off}$ is secondary to the effects of slit alignment errors. We note that the values of these offsets reflect {\it both} the pointing accuracy of the telescope and the ability to place the RC center in the slit for the reasons stated above.  The relation between a single slit alignment error and the resulting $x_{off}$ and $V^x_{off}$ is nontrivial, as it depends on the direction of the galaxy displacement relative to the galaxy major axis and to the long axis of the slit\footnote{In particular, we note that a displacement along the length of the slit will not change $x_{off}$ and $V^x_{off}$, irrespective of $\Delta PA$. These offsets thus do not reflect the absolute position of the galaxy in slit coordinates.}. For the sample as a whole, however, we expect the distributions of $x_{off}$ and $V^x_{off}$ to reflect the value of $\mathbf{r_{off}}$ corresponding to the long-slit observations. 

  We also look for trends between $|x_{off}|$, $|V^x_{off}|$ and the galaxy properties used in constructing the subsamples of Table~\ref{KS}. As an example, Fig.~\ref{xoffdist} plots $|x_{off}|$ and $|V^x_{off}|$ for all sample galaxies as a function of $D$, a property that correlates well with $\alpha_m$ (Figs.~\ref{sam_trends}a and \ref{rin}). There is some evidence for a larger median $|x_{off}|$ in nearby systems, likely because of their larger size relative to the slit.
The trend is weak, however, and is not evident in $|V^x_{off}|$. We find no correlation between  $x_{off}$, $V^x_{off}$ and other galaxy properties, or in the value of $\alpha_m$ obtained (see Figs.~\ref{sam_trends}g and \ref{sam_trends}h). We therefore use the distributions of $x_{off}$ and $V^x_{off}$ for all sample galaxies to select the simulation inputs, noting that this results in a conservative estimate of $\mathbf{r_{off}}$ for samples restricted to nearby galaxies. We use the same input parameters for subsamples 6 and 8 as well, as only the outlying $|x_{off}|$ and $|V^x_{off}|$ are pruned therein. 

 We choose $\Delta PA$, $\mathbf{r_{off}}$ and $V_{off}$ as follows. For each galaxy, we conservatively assign $\Delta PA$ according to the $i$-dependent PA error relation derived from ellipse fits to the SFI++ I-band data (H99) shown in Fig.~\ref{errors}d, such that $\Delta PA(i)$ is a Gaussian random deviate with mean  $\overline{\Delta PA(i)}$ and standard deviation $\sigma_{PA}(i)$. Here $\Delta PA(i)$ may either increase or decrease the slit PA relative to the galaxy major axis. We assume that $\mathbf{r_{off}}$ is in a random direction on the sky, and choose it as a Gaussian random deviate with standard deviation $\sigma_{roff}$. We select $\sigma_{roff}$ so that the distributions of $x_{off}$ and $V^x_{off}$ recovered during the analysis of the simulated observations resemble those obtained for the sample data, and adopt  $\sigma_{roff}=1.3^{\prime \prime}$ to this end. Fig.~\ref{errors} compares typical distributions of $x_{off}$ and $V^x_{off}$ for the sample data and the corresponding simulated observations with this choice. The value of $V_{off}$ should have no effect on the RCs themselves, as a constant velocity offset is subtracted from the raw spectrum when the kinematic center is determined. For concreteness, we choose $V_{off}$ as a Gaussian random deviate with standard deviation $\sigma_{voff}=5\,\,\rm{km\,s^{-1}}$.

To compute the raw spectrum, we define $(x, y)$ as slit coordinates with $x$ along the long axis and $(x^{\prime}, y^{\prime})$ as galaxy coordinates with $x^{\prime}$ along the major axis. The distribution of measured velocities in each pixel of the spectrograph is then obtained by summing the parts of the projected galaxy disk that fall within the slit:
\begin{equation}
P(x,V) = \Sigma_{x}\Sigma_{y}\,\textrm{H(} r_{opt} \textrm{)}\delta[V-V_{||}(x^{\prime},y^{\prime})],
\label{extract}
\end{equation}
where H($r_{opt}$)=1 for $|r|\leq r_{opt}$ and 0 otherwise, $\delta$ is the Dirac-delta function, and $V_{||}(x^{\prime},y^{\prime})$ is the component of $V^{\alpha_{int}}_h(x^{\prime},y^{\prime})$ along the line of sight. The sum in $x$ is over half a seeing disk and that in $y$ is over the width of the slit. The grid spacing in slit coordinates is 0.\arcsec 1 in all simulations.
We then construct the spectral profile by smoothing with a Hanning window and taking every third pixel as an RC point, yielding an effective resolution of 2\arcsec\ . A turbulent term is added to each point such that the distribution of $\delta V$ in the simulations matches that for the data (Fig.~\ref{errors}c). At this stage each profile is examined, and only those of quality comparable to or better than the sample RCs are further processed. For the subsample distributions of $V_{rot}$, $\delta V$, $D$, and $i$ from which the galaxy parameters are chosen, the simulated RCs are further processed $\sim 90\%$ of the time. These RCs are then folded and $\alpha_m$ for each simulated galaxy is derived using the same algorithms as described in \S~\ref{inversion}.

\section{Comparing Simulations and Data}
\label{results}

The large number of galaxies in our sample allows for a quantitative comparison of $\alpha_m$ obtained for the data with those expected from long-slit observations for a population with a specified $\alpha_{int}$.  As each simulated spectral profile must be examined during the data analysis stage, there is a trade-off between the number of galaxies one simulates for comparison with sample data and the reliability with which that comparison can be made. We find that a simulated population twice the size of the subsample in question is optimal in this case, and adopt this prescription throughout. 

To assess the level of agreement between a given subsample and a corresponding simulation, we compute a Kolmogorov-Smirnov (K-S) statistic. The K-S test evaluates the probability that the distributions of $\alpha_m$ for the sample and simulated galaxies are selected from the same parent distribution. We denote the P-value of a K-S test between the distribution of $\alpha_m$ for a subsample $k$ (where $k=0-8$) and that output from a simulation with a given $\alpha_{int}$ as $P(k,\alpha_{int})$. We find that  $P(k,\alpha_{int})$ may vary by factors of $\sim2$ when the simulations for a given subsample are repeated.
 We compare our data with simulation outputs for $\alpha_{int}=1$ in \S~\ref{int1}, and for simulated halos with different values of $\alpha_{int}$ in \S~\ref{vary}. 

\begin{deluxetable}{crrrr}
\tablecaption{K-S Test Results for Simulations with Different $\alpha_{int}$ \label{KS2}} 
\tablewidth{0pt}
\tablehead{
\colhead{Subsample} & \colhead{$P(k,1/2)$} & \colhead{$P(k,1)$} &  \colhead{$P(k,5/4)$} &  \colhead{$P(k,3/2)$} \\
\colhead{(1)} & \colhead{(2)} & \colhead{(3)} & \colhead{(4)} & \colhead{(5)}
}
\startdata
0 & 0.05 & 0.4 & 0.001 & 1E-16 \\
1 & 0.2 & 0.2 & 0.2 & 6E-15\\
2 & 0.08 & 0.5 & 0.002 & 5E-12\\
3 & 0.3 & 0.07 & 0.006 & 3E-13\\
4 & 0.04 & 0.2 & 0.3 & 3E-5\\
5 & 0.7 & 0.08 & 0.05 & 2E-10\\
6 & 0.08 & 0.9 & 0.6 & 1E-6\\
7 & 0.05 & 0.3 & 0.1 & 4E-8\\
8 & 0.005 & 0.9 & 0.4 & 1E-5\\
\enddata
\tablecomments{(1) Subsample name. (2) -- (5) P-value $P(k,\alpha_{int})$ of a K-S test between subsample $k$ and a simulation with inner slope $\alpha_{int}$.}
\end{deluxetable}

\subsection{Simulations with $\alpha_{int}=1$}
\label{int1}

\begin{figure*}
\epsscale{0.9}
\plotone{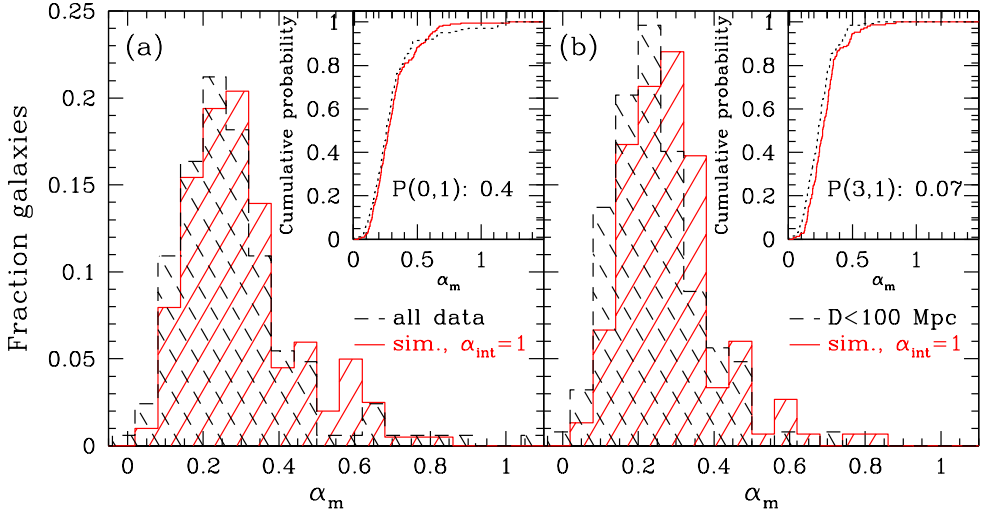}
\caption{Comparison between actual (dashed lines) and simulated (solid red lines) $\alpha_m$ for a) all sample galaxies, and b) subsample 3, where $D<100$ Mpc. In both a) and b), the main panel shows the distributions of $\alpha_m$, and the corresponding cumulative probability distributions are inset. The P-value $P(k,1)$ of a K-S test is also indicated. 
 \label{simint1}}
\end{figure*}

\begin{figure}
\epsscale{1.25}
\plotone{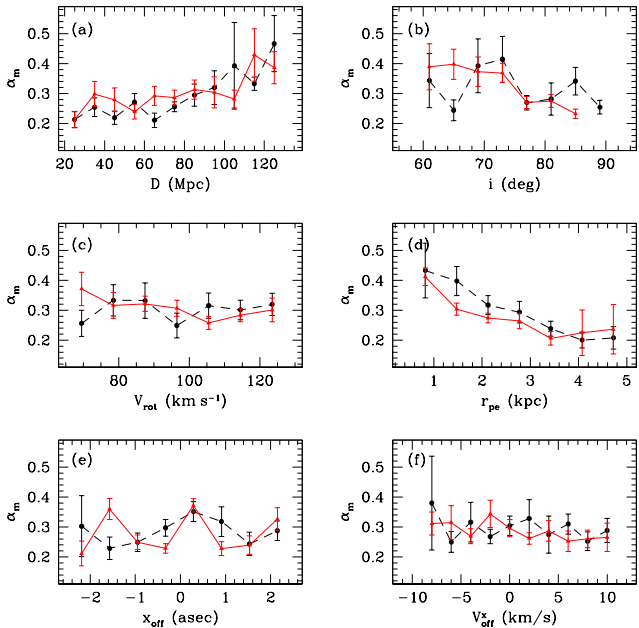}
\caption{Mean value of $\alpha_m$ for a simulated population of halos with $\alpha_{int}=1$ as a function of various galaxy properties (solid red lines), superimposed on the corresponding relations for subsample 0 (dashed lines). Plot details are as in Fig.~\ref{sam_trends}. 
 \label{sim_trends}}
\end{figure}

We simulate long-slit observations of a series of dwarf populations, with $\alpha_{int}=1$ and  observed properties that match those in each of the 8 subsamples defined in \S~\ref{sample}. For each subsample $k$, $P(k,1)$ is listed in col. 3 of Table~\ref{KS2}.
Fig.~\ref{simint1} compares the distributions of $\alpha_m$ obtained for subsamples 0 (all galaxies) and 3 ($D<100$~Mpc) with those obtained for simulated populations with the same statistical properties as each of these subsamples. Qualitatively, there is reasonable agreement between the simulated and actual $\alpha_m$ for both subsamples.  The simulations also recover the same trends in $\alpha_m$ with simulated galaxy properties as found for the sample galaxies (see Fig.~\ref{sam_trends}): we show the results for all sample galaxies in Fig.~\ref{sim_trends}, to demonstrate the correspondence between the simulations and the data over the broadest range of parameter space.

In Fig.~\ref{simint1}, the subsample containing all the galaxies (subsample 0) is better recovered than that containing only the galaxies with $D<100$~Mpc; the latter simulation produces a longer tail towards large $\alpha_m$ than do the data, and underpredicts the number of galaxies with small $\alpha_m$. 
Quantitatively, the K-S statistics computed from the cumulative probability distributions inset in Fig.~\ref{simint1} differ substantially, with $P(3,1)=0.07$ and $P(0,1)=0.4$. The pairs of distributions in Fig.~\ref{simint1} thus provide an indication of the sensitivity the K-S test to differences between input distributions of $\alpha_m$ given the sample characteristics.

There is no clear correlation between the selection criteria for each subsample $k$ and the value $P(k,1)$ returned in Table~\ref{KS2}. Subsample 4, for example, has both the lowest median $\alpha_m \pm \Delta \alpha_m$ and the most stringent $D$ cutoff, but $P(4,1)$ lies in between those of the distributions shown in Fig.~\ref{simint1}. A direct comparison between $P(k,1)$ for the different subsamples is somewhat muddled, however, by the different numbers of galaxies therein: in particular, the subsamples with the strictest selection criteria are the smallest, and thus more extreme differences between the actual and simulated distributions of $\alpha_m$ are required to produce a low $P(k,\alpha_{int})$ than in larger subsamples. Indeed, the two subsamples with the lowest $P(k,1)$ are subsamples 3 ($D<100$~Mpc) and 5 ($M_I<-20.5$), which have the largest number of galaxies other than subsample 0.

 Even for the larger subsamples with $P(k,1)$ at the low end of the range $0.07\lesssim P(k,1) \lesssim 0.9$, there is little evidence for a statistically significant difference between the distributions of $\alpha_m$ for the sample and corresponding simulations.  In other words,  the $\alpha_m$ values obtained for the sample RCs are consistent with a scenario in which low-mass galaxies have intrinsically cuspy halos with $\alpha_{int}=1$. We note, however, that our data {\it do not} require that low-mass galaxies have cusps instead of cores.

\begin{figure}
\epsscale{1.2}
\plotone{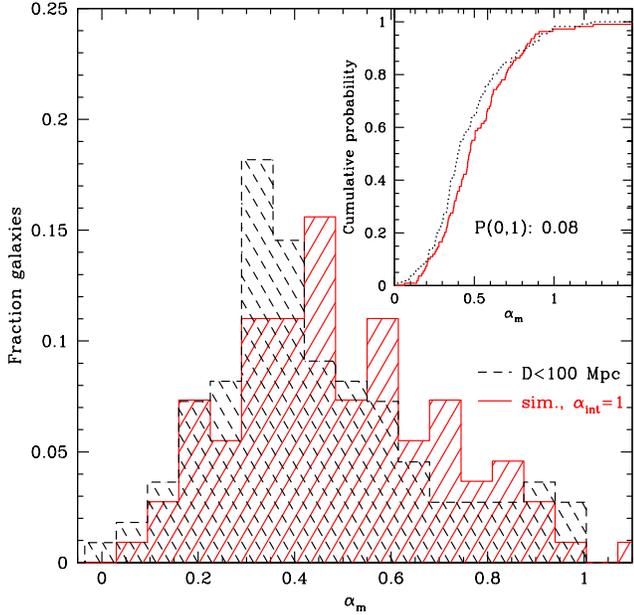}
\caption{Same as in Fig.~\ref{simint1}a, but with all points $r_i <$~1.5\arcsec\ excluded in both the sample and simulation analyses. 
 \label{r15}}
\end{figure}

 It is important to verify that the agreement between the distribution of $\alpha_m$ output from the simulations with that for the data is robust to changes in our measurement technique. To illustrate this, we note that the $r_i$ values within a seeing disk of $r=0$ are highly uncertain because of potential RC folding errors, in addition to the large relative errors in $V(r_i)$ noted earlier. Accordingly, we re-measure $\alpha_m$ for the galaxies in subsample 3, this time discarding all RC points with $r_i<1.5$\arcsec\ . 
We then simulate a galaxy population and analyze the spectral profiles obtained with this same adjustment to our technique. 
 The results of this exercise are shown in Fig.~\ref{r15}.
  The median values of $\alpha_m$ for the simulations and the data are larger than in Fig.~\ref{simint1}b: the method by which $\alpha_m$ is measured clearly affects the values obtained. As before, however, the distribution of simulated $\alpha_m$ resembles that for the data. 
Similarly, we verify that changes in the median input $V_{200}$ of 50 $\rm{km\,s^{-1}}$ and corresponding adjustments to $c$ do not significantly alter the distribution of $\alpha_m$ output from the simulations.

In summary, we can recover the shallow distribution of $\alpha_m$ found in the data by simulating long-slit observations of cuspy ($\alpha_{int}=1$) dwarf halos. We find a range $0.07\lesssim P(k,1) \lesssim 0.9$ among the subsamples, with the largest corresponding to subsample 0 and the smallest to subsample 3. The simulation results are robust to changes in the data analysis technique and to small variations in the halo shapes input in the simulation. 
We thus find that while individual galaxies show strong departures from $\alpha_{int}=1$, the sample $\alpha_m$ are {\it statistically consistent} with $\alpha_{int}=1$ halos, when observing and data processing biases are taken into account.

\subsection{Simulations with $\alpha_{int} \ne 1$}
\label{vary}
 
\begin{figure*}
\epsscale{0.8}
\plotone{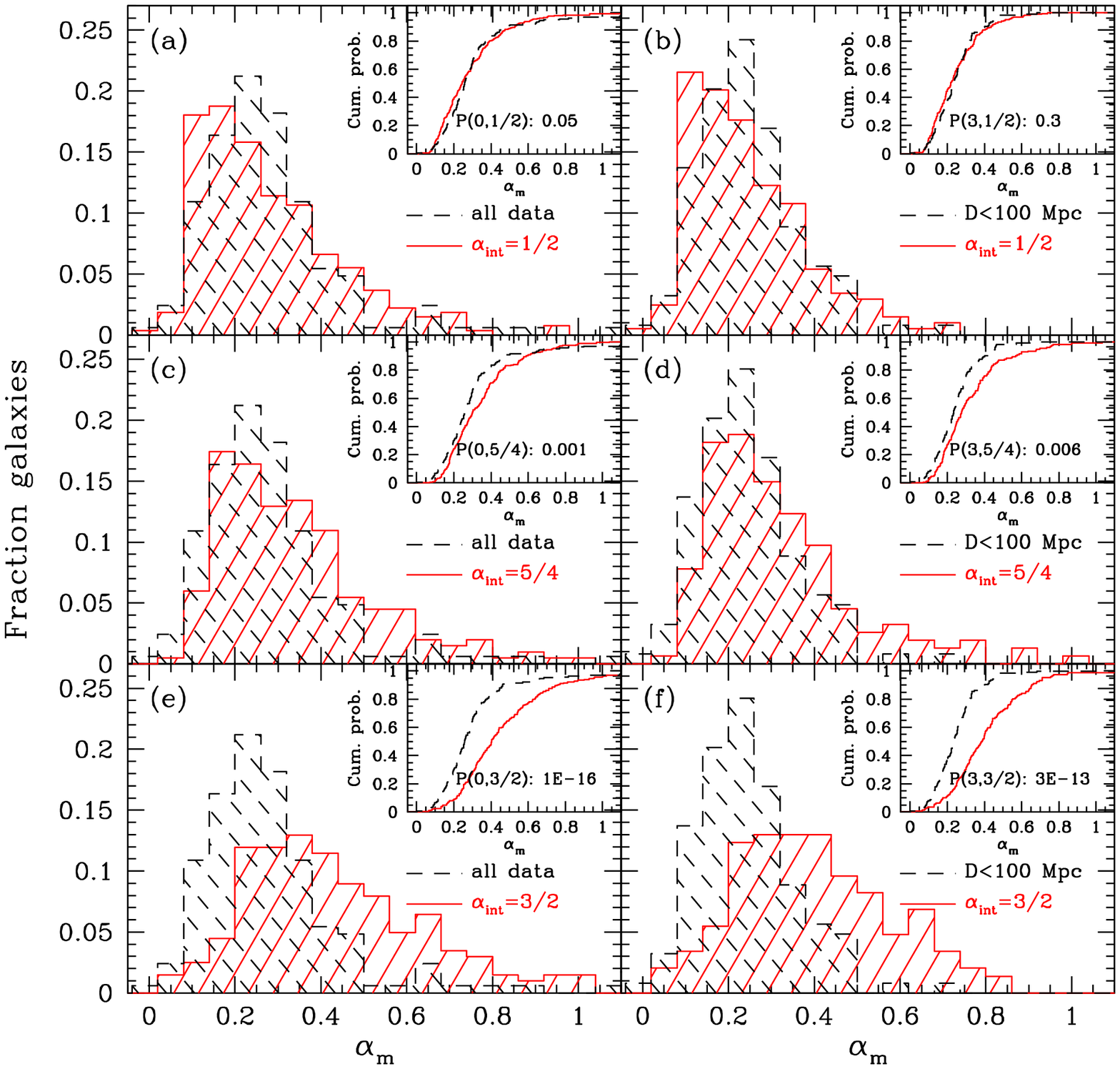}
\caption{Comparison between distributions of $\alpha_m$ for subsample 0 (all galaxies; left column) and for subsample 3 ($D<100$~Mpc; right panels) and corresponding simulation outputs for $\alpha_{int}=1/2$ (a, b),  $\alpha_{int}=5/4$ (c, d), and  $\alpha_{int}=3/2$ (e, f). In each panel, the plot details are the same as in Fig.~\ref{simint1}. 
 \label{simintne1}}
\end{figure*}

For each subsample of galaxies, we carry out a series of simulations for which $\alpha_{int} \ne 1$. 
 The main results are given in Table~\ref{KS2}, where we list $P(k,1/2)$ in col. 2, $P(k,5/4)$ in col. 4, and  $P(k,3/2)$ in col. 5. 
 In Fig.~\ref{simintne1}, we plot the distributions of $\alpha_m$ for all 165 galaxies (subsample 0; left column) as well as for galaxies with $D<100$~Mpc (subsample 3; right column) and the corresponding simulation outputs for populations with $\alpha_{int}=1/2$ (panels a and b), $5/4$ (panels c and d), and $3/2$ (panels e and f).  Qualitatively, the distribution of $\alpha_m$ for simulated halos with $\alpha_{int}=1/2$ or 5/4 for both subsamples 0 and 3 are again similar to the distributions found in the data. The histograms corresponding to $\alpha_{int}=1/2$ are slightly shallower than that for the data, however, and for $\alpha_{int}=5/4$ there is a larger fraction of simulated systems in the tail extending to $\alpha_m \sim 1$. There is also a larger variation in $P(k,1/2)$ and $P(k,5/4)$ among subsamples than for the $\alpha_{int}=1$ case, with $0.005\,\le\,P(k;1/2)\,\le\,0.7$ and $0.001\,\le\,P(k;5/4)\,\le\,0.6$.

 In general, there is a good correspondence between $\alpha_m$ for the simulated populations and for the subsample data when $\alpha_{int}=1/2$: the larger subsamples have $P(k,1/2)>0.1$ and all but one have $P(k,1/2)>0.01$. In particular, the values of $P(k,1/2)$ for subsamples 3 and 5  are larger than the corresponding $P(k,1)$.
This is not the case for all subsamples with low median $\alpha_m$, however: both subsamples 4 and 7 have lower $P(k,1/2)$ than $P(k,1)$. Also in contrast to the simulation results for $\alpha_{int}=1$, the smallest subsamples (2, 4 and 8) are the ones with the most extreme differences between the simulated and actual distributions of $\alpha_m$.

 The simulations with $\alpha_{int}=5/4$ tend not to recover the observed distribution of $\alpha_m$ as well as either the $\alpha_{int}=1/2$ or $\alpha_{int}=1$ cases. This is evident in the cumulative distributions inset in Figs.~\ref{simintne1}c~and~d, in which the simulations for both subsamples 0 and 3 underpredict the fraction of halos with $0.3 \lesssim \alpha_m \lesssim 0.5$. While $P(k,5/4)\geq 0.1$ for over half of the subsamples, the latter tend to be the smallest ones considered, and thus more extreme differences between the simulated and actual $\alpha_m$ are required to lower $P(k,5/4)$ than for the larger samples.  There is no clear trend between the median $\alpha_m$, selection criteria or size of a given subsample and $P(k,5/4)$, however. We thus find that long-slit observations of halos with $\alpha_{int}=5/4$ yield distributions of $\alpha_m$ that are only marginally consistent with the sample galaxies. Sample sizes larger than those examined here are needed to further investigate this case.

For $\alpha_{int}=3/2$, our mock long-slit observations of galaxy models (Figs.~\ref{simintne1}c~and~d) yield a distribution of $\alpha_m$ that is clearly discrepant from that found in the data. We do find some small $\alpha_m$ despite the steep input $\alpha_{int}$; however, most of the values are too large to be consistent with the core-like distributions in the data. As such, $P(k,3/2)<<1$ for all subsamples, and the hypothesis that the distributions of $\alpha_m$ for the data and $\alpha_{int}=3/2$ simulations stem from the same parent distribution is strongly rejected. 

\section{Discussion and Summary}
\label{discuss}

Under the assumptions of a minimum disk and spherical symmetry, we have measured the inner halo shapes $\alpha_m$ for a sample of 165 dwarf galaxies from long-slit optical spectra. We find a median value $\alpha_m = 0.26 \pm 0.07$ for all sample RCs, and similar values for various subsamples of the data (Table~\ref{KS}). Irrespective of the subsample in question, the distribution of $\alpha_m$ is significantly shallower than the cusps ($\alpha_{int}$) predicted by the CDM paradigm (e.g. Figs.~\ref{singlehist},~\ref{rin}).  Our results resemble those obtained in other studies in which similar analyses were performed, albeit with smaller samples (e.g. dB01; dB03; S03): it thus seems that in general, halo shapes inferred via the direct inversion of long-slit spectra (eq.~(\ref{invers})) exhibit the cusp/core problem. The variation of $\alpha_m$ with $r_{in}$ (Fig.~\ref{rin}) clearly demonstrates the need for high-resolution observations in determining $\alpha_m$ from RC shapes.

 To elucidate the relationship between $\alpha_m$ and various intrinsic inner slopes $\alpha_{int}$ that are consistent with the CDM paradigm, we simulate long-slit observations of model galaxies with the same global properties as each of our subsamples.
The clear dependence of $\alpha_m$ on $D$ (Figs.~\ref{singlehist},~\ref{sam_trends}a,~\ref{rin}) as well as on the measurement technique (Fig.~\ref{r15}), and the wide range of $\alpha_m$ measured for model galaxies in these simulations (Fig.~\ref{simintne1}) underscore the importance of simulations in assessing the nature of the discrepancy between $\alpha_m$ and $\alpha_{int}$.  

Our simulations show that populations of dwarf galaxies with $\alpha_{int}=$ 1/2 or 1 observed with a long slit yield distributions of $\alpha_m$ that are broadly consistent with those obtained for the sample RCs (Figs.~\ref{simint1},~\ref{simintne1}a--b). These simulations also recover trends in $\alpha_m$ with other galaxy properties (Fig.~\ref{sim_trends}), suggesting that they capture the essential processes that govern the value of inner halo slope measured. The distribution of $\alpha_m$ in our sample is therefore consistent with halo profiles that are shallower than or comparable to the NFW case: the discrepancy between $\alpha_m$ and $\alpha_{int} \sim 1$ in our sample is reconciled when the impacts of observing and data processing techniques are considered.

 We emphasize that this result {\it does not} imply that the halos in our sample actually have cusps with $\alpha_{int} \sim 1$: the available data are merely {\it statistically consistent} with them. Indeed, it has been repeatedly demonstrated (e.g. dB03; S03) that corelike distributions of measured inner slopes such as those found here are also consistent with a variety of intrinsically corelike halo shapes. Moreover, our findings clearly do not address the nature of the cusp/core problem obtained from other types of analyses (see \S~\ref{intro}). However, they do demonstrate that, for a large sample of dwarf galaxies, the distinction between intrinsic $\alpha_{int}=1$ cusps and cores cannot be made by measuring $\alpha_m$ under assumptions of minimal disks and spherical symmetry. 

  We find that simulated observations of model galaxies with $\alpha_{int}=5/4$ are only marginally consistent with the data in our sample (Fig.~\ref{simintne1}c--d). It is therefore unclear whether observing and data processing techniques are sufficient to reconcile $\alpha_m$ with this case, and the cusp/core problem may remain should CDM simulations converge to halos with $\alpha_{int} \sim 1.3$ on galactic scales (e.g. Navarro et al. 2003). A sample size larger than the one considered here may offer firmer conclusions for the $\alpha_{int}=5/4$ case. More sophisticated simulations that include the effects discussed by Rhee et al. (2004) should also be considered in this case. The hypothesis that dwarf galaxies have $\alpha_{int}=3/2$ is strongly rejected by our data (Figs.~\ref{simintne1}e--f), and we do not expect that adding complexity to our models will change this result.

 For dark matter-dominated systems such as the ones considered here, there is typically little change in $\alpha_m$ when the luminous components are included in the analysis (dB01, dB03, S03). In addition, our simulations reproduce trends in $\alpha_m$ with various galaxy properties found in the data (Fig.~\ref{sim_trends}), indicating that they do capture the processes that govern the value of $\alpha_m$ obtained. For these reasons, we expect that the assumption of a minimum disk does not significantly affect our results. Indeed, the extra parameters introduced in mass model fits by including luminous components renders the $\alpha_{int}=1$ case prohibitively difficult to rule out, even in the absence of the RC uncertainties addressed here (S03). It is therefore unclear whether the inner regions of long-slit H$\alpha$ RCs should be used to address the cusp/core problem at all: studies that exploit the RC shape near $r_{opt}$ rather than at $r \sim 0$ should provide more robust constraints on the halo profile (e.g. Salucci 2001).

 High-resolution two-dimensional analyses of spiral galaxy velocity fields are immune to many uncertainties that plague those from long-slit spectra, and may therefore be able to distinguish various $\alpha_{int}$ from the measured $\alpha_m$. 
These observations are underway: Bolatto et al. (2004) and Blais-Ouellette et al. (2004) present RCs for a total of 11 systems derived from H$\alpha$ + synthesis CO maps and Fabry-Perot H$\alpha$ maps, respectively. 
The 5 galaxies of Bolatto et al. have different halo structures ranging from $0.3 \lesssim \alpha_m \lesssim 1.1$
The data from Blais-Ouellette et al. are thoroughly analyzed by Dutton et al. (2003), who run an extensive set of mass models with contributions from stellar and gaseous disks and allow for halo shapes considerably more complex than those considered here.
They find that $\alpha_{int} = 0$ provides a better fit to
the data than halos with $\alpha_{int} = 1$ for a range of model
parameters.

provides the best fit for all but one of their 6 RCs, but cannot conclusively rule out  $\alpha_{int} \ge 1$ systems.

 As the quality of spiral galaxy kinematics derived from observations increases, the presence of baryons in dark matter halos cannot be ignored in CDM paradigm tests. The discrepancy between cusp and core in mass models increases if the disk contracts adiabatically in the halo (dB01; Dutton et al. 2003; S03). However, the implicit assumption that the initial angular momentum distributions of the baryons and the dark matter are the same has been called into question (e.g. Bullock et al. 2001; van den Bosch et al. 2001), and Gnedin et al. (2004) find that adiabatic contraction over-estimates the true degree of halo contraction.  More dramatic gasdynamical effects, such as star formation and feedback during disk formation, also need to be considered.  The impact of central starbursts during disk formation was simulated by Mo \& Mao (2004), who find lower c and $\alpha_{int}$ relative to those of collisionless halo shapes. As such, it is perhaps not surprising that the halo shapes inferred in regions with strong baryonic contributions to the mass density do not match the profile shapes of the collisionless, precollapse halos discussed here (Weiner et al. 2001, Binney 2004).
Indeed, the discrepancy between CDM theory and observations is largest at low redshift, where gasdynamical effects are most important: the role of these effects in galaxies and cluster -- sized halos needs to be assessed before a CDM crisis is advocated.

\acknowledgements
This research has been partially funded by NSF grant AST-0307396 to RG. We thank the referee of this paper, whose comments improved its content and clarity.
This research has made use of the NASA/IPAC Extragalactic Database which is operated by the Jet Propulsion Laboratory, California Institute of Technology, under contract with the National Aeronautics and Space Administration.
This publication makes use of data products from the Two Micron All Sky Survey, which is a joint project of the University of Massachusetts and the Infrared Processing and Analysis Center/California Institute of Technology, funded by the National Aeronautics and Space Administration and the National Science Foundation.

\end{document}